\documentclass[10pt,twocolumn,oneside,final]{IEEEtran}
\IEEEoverridecommandlockouts
\usepackage{cite}
\usepackage{graphicx}
\usepackage{amsmath}
\usepackage{times}
\usepackage{latexsym}
\usepackage{bm}
\usepackage{amssymb}
\usepackage[center]{caption2}
\usepackage{array}
\usepackage{fancyhdr}
\usepackage{cite,graphicx,amsmath,amssymb}
\usepackage{psfrag}
\usepackage{multirow}
\usepackage{array}
\usepackage{subfigure}
\usepackage{stfloats}
\usepackage{url}
\usepackage{nicefrac}
\usepackage{amssymb}
\usepackage{amsmath}
\usepackage{amsfonts}
\usepackage{amssymb}
\usepackage{float}
\usepackage{kbordermatrix}

\ifCLASSINFOpdf

\else

\fi

\usepackage[dvipsnames,usenames]{color}
\usepackage[usenames,dvipsnames]{color}

    \def\Complex{{\rm\rule[.23ex]{.03em}{1.1ex}\kern-.3em{C}}}

    \newcommand{\be}{\begin{equation}} \newcommand{\ee}{\end{equation}}
    \newcommand{\bea}{\begin{eqnarray}} \newcommand{\eea}{\end{eqnarray}}
    \newcommand{\benum}{\begin{enumerate}} \newcommand{\eenum}{\end{enumerate}}

    \newcommand{\qG}{{\bf G}}

    \newcommand{\tr}{{\sf tr}}

\newtheorem{theorem}{Theorem}

\newtheorem{alg}{Algorithm}
\newtheorem{definition}{Definition}

\newtheorem{example}{Example}

\begin{document}
\title{Secure Transmission with Large Numbers of Antennas and Finite Alphabet Inputs}

\author{Yongpeng Wu, \IEEEmembership{Senior Member, IEEE}, Jun-Bo Wang, \IEEEmembership{Member, IEEE}, Jue Wang, \IEEEmembership{Member, IEEE} \\
Robert Schober, \IEEEmembership{Fellow, IEEE}, and Chengshan Xiao, \IEEEmembership{Fellow, IEEE}


\thanks{Part of this paper was submitted to IEEE Globecom 2017.}

\thanks{Y. Wu is with Institute for Communications Engineering,  Technical University of Munich,
Theresienstrasse 90, D-80333 Munich, Germany (Email:yongpeng.wu2016@gmail.com).}

\thanks{Y. Wu and R. Schober are with the Institute for Digital Communications, University Erlangen-N$\ddot{u}$rnberg,
Cauerstrasse 7, D-91058 Erlangen, Germany (Email: yongpeng.wu@fau.de; robert.schober@fau.de).}

\thanks{J.-B. Wang is with the National Mobile Communications Research Laboratory,
Southeast University, Nanjing, 210096, P. R. China (Email: jbwang@seu.edu.cn).}

\thanks{J. Wang is with School of Electronics and Information, Nantong University, Nantong, 226000, P. R. China (Email: wangjue@ntu.edu).}

\thanks{C. Xiao is with the Department of Electrical and Computer Engineering,
Missouri University of Science and Technology, Rolla, MO 65409, USA (Email: xiaoc@mst.edu). }

}
\maketitle

\begin{abstract}
In this paper, we investigate  secure transmission over the large-scale
multiple-antenna wiretap channel with finite alphabet inputs.
First, we investigate the case where instantaneous channel state information (CSI)
of the eavesdropper is known at the transmitter. We show analytically that a generalized singular value decomposition (GSVD)
based design, which is optimal for Gaussian inputs,
may exhibit a severe performance loss for finite alphabet inputs in the high signal-to-noise ratio (SNR) regime.
In light of this, we propose a novel Per-Group-GSVD (PG-GSVD) design
which can effectively compensate the performance loss caused by the GSVD
design. More importantly,  the computational complexity of the PG-GSVD design  is
 by orders of magnitude lower than that of the existing
design for finite alphabet inputs in \cite{Wu2012TVT} while the resulting performance loss is minimal.
Then, we extend the PG-GSVD design to the case where only statistical CSI of the eavesdropper
is available at the transmitter.
Numerical results  indicate that the proposed PG-GSVD design can be efficiently implemented
in large-scale multiple-antenna systems and achieves significant performance gains compared to the GSVD design.
\end{abstract}


\section{Introduction}
Security is a critical issue for future 5G
wireless networks. In today's systems, the security provisioning relies on bit-level cryptographic mechanisms
and associated processing techniques at various stages of the data protocol stack. However, these solutions have
severe drawbacks and many weaknesses of
standardized protection mechanisms for public wireless networks  are well known;
although enhanced ciphering and authentication protocols
exist, they impose  severe constraints and high additional costs for the users of public wireless networks \cite{Yang2015CM}.
Therefore, new security approaches based on information theoretical considerations
have been proposed and are collectively referred to as physical layer security  \cite{Wu2012TVT,Zhu2014,ZhuJ2016TWC,Yang2015CM,Wyner1975BSTJ,Shafiee2007ISIT,Tekin2008TIT,Khisti2010TIT,Khisti2010TIT_2,Li2011TWC,Gursoy2012TCom,Zhu2014TSP,Liu2014TSP,Rezki2014TWC,Li2015commletter,Chen2015CM,Chen2015TWC,Zhu2016TWC,Li2016TCOM,Wang2016TSP,Wang2016TSP_2,Zeng2016TWC,Wu2016TIT,Zou2016PIEEE,Zou2017TWC,Bashar2011Cletter,Bashar2012TCom,Aghdam2016}.

Most existing work on physical layer security assumes that the input signals are Gaussian distributed.
Although the Gaussian codebook has been proved to achieve the secrecy capacity of the Gaussian wiretap channel \cite{Khisti2010TIT_2},
the signals employed in practical communication systems are non-Gaussian and are often drawn from
 discrete constellations \cite{Lozano2006TIT,Xiao2011TSP,Wu2012TWC,Wu2013TCOM}. For the multiple-input, multiple-output, multiple antenna eavesdropper (MIMOME) wiretap
channel with perfect channel state information (CSI) of both the desired user and the eavesdropper
at the transmitter, a generalized singular value decomposition (GSVD) based
precoding design was proposed to decouple the corresponding wiretap channel into  independent parallel subchannels \cite{Bashar2012TCom}.
Then, the optimal power allocation policy across these subchannels was obtained by an iterative algorithm.
However, the simulation results in \cite{Wu2012TVT} revealed that for finite alphabet inputs, the GSVD design is suboptimal.
In fact, the iterative algorithm in \cite{Wu2012TVT} can significantly improve the
secrecy rate by directly optimizing the precoder matrix.
Furthermore, for finite alphabet inputs,  both the receiver and the
eavesdropper may accurately decode a transmitted message
if the transmit power is sufficiently high.
Therefore, the optimal precoder may not exploit
the maximum available transmit power to
maximize
the secrecy rate in the high signal-to-noise ratio (SNR) regime \cite{Bashar2011Cletter,Bashar2012TCom,Wu2012TVT}.
Instead, it may be beneficial for the transmitter to use some of the available transmit
power to inject  artificial noise (AN)
to interfere the decoding process of the eavesdropper.
It has been shown in \cite{Bashar2011Cletter,Wu2012TVT} that injection of AN
improves the secrecy rate
for scenarios
where the transmitter has only statistical CSI of the eavesdropper.
Very recently, for the case when imperfect CSI of the eavesdropper is available at the transmitter,
a secure transmission  scheme  was proposed in \cite{Aghdam2016}
based on the joint design of the transmit precoder matrix to improve the achievable rate
of the desired user and the AN generation scheme to
degrade the achievable rate of the eavesdropper.
However, the computational complexities  of the algorithms in
 \cite{Wu2012TVT} and \cite{Aghdam2016} scale
exponentially with the number of transmit antennas.
Therefore, the algorithms in \cite{Wu2012TVT,Aghdam2016} become intractable
even for a moderate number of transmit antennas (e.g., eight).

In this paper, we investigate the secure transmission design
for the large-scale MIMOME wiretap channel with finite alphabet inputs.
The contributions of our paper are summarized as follows:

\begin{enumerate}

\item For scenarios where the instantaneous CSI of the eavesdropper is available
at the transmitter, we derive an upper bound on the secrecy rate for finite alphabet inputs in the high SNR regime
when the GSVD design is employed.
The derived expression shows that,  when $N_t > N_1$, in the high SNR regime, the GSVD design will result
in at least $(N_t - N_1) \log M$ b/s/Hz rate loss compared to the maximal rate
for the MIMOME wiretap channel, where $N_t $, $N_1$, and $M$ denote
the number of transmit antennas, the rank of the intended receiver's channel, and the size of the input signal constellation set,
respectively.

\item To tackle this issue, we propose a novel Per-Group-GSVD (PG-GSVD)
design, which pairs different subchannels into different groups based on the GSVD structure.
We prove that the proposed PG-GSVD design can eliminate the performance loss of the GSVD design
with an order  of magnitude lower computational complexity than the design in \cite{Wu2012TVT}. Accordingly, we propose an iterative algorithm based on
the gradient descent method to optimize the secrecy rate.

\item For the scenarios where only  statistical CSI of the eavesdropper is available
at the transmitter, we derive an achievable ergodic secrecy rate expression
by invoking an upper bound on the average
mutual information in fading multiple-input, multiple-output (MIMO) channels with  finite alphabet inputs. Based on this, we extend the PG-GSVD
design to the case where the transmitter has only statistical CSI of the eavesdropper.

\item By exploiting the low rank property of the transmit correlation
matrices of massive MIMO channels \cite{Adhikary2013TIT,Yin2014JSTSP,Wu2016TIT}, we derive
a condition
that the proposed PG-GSVD design
for statistical CSI of the eavesdropper should satisfy to achieve the maximal secrecy rate for the MIMOME wiretap channel
with finite alphabet inputs  in the high SNR regime. For the cases where this
condition does not hold, we propose an AN generation scheme that can further
increase the secrecy rate performance.

\item Simulation results illustrate that the proposed designs are well suited for large-scale MIMO wiretap channels and
achieve substantially higher secrecy rates than the GSVD design while requiring a much
lower computational complexity than the precoder design in \cite{Wu2012TVT}.

\end{enumerate}

\emph{Notation:}  Vectors are denoted by lower-case bold-face letters;
matrices are denoted by upper-case bold-face letters. Superscripts $(\cdot)^{T}$, $(\cdot)^{*}$, and $(\cdot)^{H}$
stand for the matrix transpose, conjugate, and conjugate-transpose operations, respectively. We use  ${\tr}({\bf{A}})$ and ${\bf{A}}^{-1}$
to denote the trace and the
inverse of matrix $\bf{A}$, respectively.
$^{\bot}$ denotes the orthogonal complement of a subspace.
 ${\rm{diag}}\left\{\bf{b}\right\}$ denotes a diagonal matrix
with the elements of vector $\bf{b}$ on its main diagonal.
${\rm{Diag}}\left\{\bf{B}\right\}$  denotes a diagonal matrix containing in the main diagonal
the diagonal elements of matrix $\mathbf{B}$.
The $M \times M$ identity matrix is denoted
by ${\bf{I}}_M$, and the all-zero $M \times N$ matrix and the all-zero $N \times 1$ vector are denoted by $\bf{0}$.
The fields of complex numbers and real numbers are denoted
by $\mathbb{C}$ and $\mathbb{R}$, respectively. $E\left[\cdot\right]$ denotes statistical
expectation. $[\mathbf{A}]_{mn}$ denotes the element in the
$m$th row and $n$th column of matrix $\mathbf{A}$. $[\mathbf{a}]_{m}$ denotes the $m$th entry
of vector $\mathbf{a}$. We use  $\mathbf{x} \sim \mathcal{CN} \left( {\mathbf{0},{{\bf{R}}_N}} \right)$
to denote a circularly symmetric complex Gaussian vector
$\mathbf{x} \in {\mathbb{C}^{N \times 1}}$ with zero mean and covariance matrix ${\bf{R}}_N$.
$\rm{null}{\left(\mathbf{A}\right)}$ denotes the null space of matrix $\mathbf{A}$.
${\bf{e}}_i$ denotes the unit-vector with a one as the $i$th element and zeros for all other elements.

\section{System Model}
We study the MIMOME wiretap channel with a multiple-antenna transmitter (Alice), a multiple-antenna
 intended receiver (Bob), and a multiple-antenna  eavesdropper (Eve), where the corresponding numbers of antennas are denoted by
 $N_t$, $N_r$, and $N_e$, respectively. The signals
 received at Bob and Eve are denoted by ${\mathbf{y}}_b$ and ${\mathbf{y}}_e$, respectively,
and can be written as
\begin{equation}\label{main}
{\mathbf{y}}_b  =  {\mathbf{H}}_{ba} {\mathbf{Gx}}_a  + {\mathbf{n}}_b
\end{equation}
\begin{equation}\label{eave}
{\mathbf{y}}_e  =  {\mathbf{H}}_{ea} {\mathbf{Gx}}_a  + {\mathbf{n}}_e
\end{equation}
where ${\mathbf{x}}_a = [x_1, x_2, \cdots,x_{N_t}]^T \in \mathbb{C}^{N_t \times 1}$ denotes
the transmitted signal vector having zero mean and the identity matrix as
covariance matrix, and ${\mathbf{H}}_{ba} \in \mathbb{C}^{N_r \times N_t}$ and ${\mathbf{H}}_{ea} \in \mathbb{C}^{N_e \times N_t}$
denote the channel matrices between Alice and Bob and between Alice and Eve, respectively.
The complex independent identically distributed
(i.i.d.) vectors ${\mathbf{n}}_b \sim {\cal CN}(0, \; \sigma_b^2 {\bf{I}}_{N_r})$
and ${\mathbf{n}}_e \sim {\cal CN}(0, \; \sigma_e^2 {\bf{I}}_{N_e})$
represent the channel noises at Bob and Eve, respectively.
${\mathbf{G}} \in \mathbb{C}^{N_t \times N_t}$ is
a linear precoding matrix that has to be optimized
for maximization of the secrecy rate.
The precoding matrix has to satisfy
the power constraint
\begin{equation}\label{power_constraint}
{\tr}\left\{ {E\left[ {{\mathbf{Gx}}_a\mathbf{x}_a^H  {\mathbf{G}}^H  } \right]} \right\} = {\tr}\left\{ {{\mathbf{GG}}^H  } \right\} \le P.
\end{equation}

In this paper, we assume that the transmitter has
perfect instantaneous CSI of the intended receiver.
For the eavesdropper's CSI available at the transmitter, we
consider the following two cases:

\begin{enumerate}
\item The perfect
instantaneous CSI of the eavesdropper is available at the transmitter. When the transmitter has perfect
instantaneous  knowledge of the eavesdropper's channel, the achievable secrecy rate is given by  \cite{Khisti2010TIT_2}
\begin{equation}\label{secrecy_capacity}
C_{\rm{sec} } =  {\mathop {\max }\limits_{\tr\left( {{\bf{GG}}^H } \right) \le P } }  R_{\rm sec}({\mathbf{G}})
\end{equation}
\begin{equation}\label{rate}
R_{\rm sec}({\mathbf{G}}) = I \left( {{\mathbf{y}}_b ;{\mathbf{x}}_a} \right) - I \left( {{\mathbf{y}}_e ;{\mathbf{x}}_a} \right)
\end{equation}
where $I(\mathbf{y}; \mathbf{x})$ denotes the mutual information  between input $\mathbf{x}$ and output $\mathbf{y}$.

\item Only statistical CSI of the eavesdropper is available at the transmitter.
{To ensure the consistency  of the channel models for Bob and Eve,
we assume that both $\mathbf{H}_{ba}$ and $\mathbf{H}_{ea}$ are Kronecker fading MIMO channels, i.e.,
\begin{align}
{\mathbf{H}}_{ba}  = {\mathbf{\tilde{R}}}_{N_{b}}^{1/2} {\mathbf{H}}_{c} {\mathbf{\tilde{R}}}_{{N}_{t} }^{1/2} \label{Hba_model} \\
{\mathbf{H}}_{ea}  = {\mathbf{R}}_{N_{e}}^{1/2} {\mathbf{H}}_{w} {\mathbf{R}}_{N_{t} }^{1/2} \label{Hea_model}
\end{align}
where ${\mathbf{H}}_{c} \in \mathbb{C}^{N_r \times N_t}$ and ${\bf{H}}_{w} \in \mathbb{C}^{N_e \times N_t}$
are complex random matrices with independent random entries, which are distributed as $\mathcal{CN}(0,1)$.
Matrices ${\mathbf{\tilde{R}}}_{{N}_{t}}
\in \mathbb{C}^{N_t \times N_t}$ and ${\mathbf{\tilde{R}}}_{N_b} \in \mathbb{C}^{N_b \times N_b}$ denote the transmit
 and receive correlation matrices of the intended receiver, respectively,
 whereas matrices ${\mathbf{R}}_{N_{t}}
\in \mathbb{C}^{N_t \times N_t}$ and ${\mathbf{R}}_{N_e} \in \mathbb{C}^{N_e \times N_e}$ denote the transmit
 and receive correlation matrices of the eavesdropper, respectively.} When the transmitter has only statistical CSI of Eve's channel, the achievable ergodic secrecy rate is given by \cite{Wu2016TIT}
\begin{equation}\label{secrecy_erg}
\bar{C}_{\rm{sec} } =  {\mathop {\max}\limits_{\tr\left( {{\bf{GG}}^H } \right) \le P } }  \bar{R}_{\rm{sec} }({\mathbf{G}})
\end{equation}
\begin{equation}\label{rate_erg}
\bar{R}_{\rm{sec} }({\mathbf{G}}) = I \left( {{\mathbf{y}}_{b} ;{\mathbf{x}}_a} \right) - E\left[I \left( {{\mathbf{y}}_{e} ;{\mathbf{x}}_a} \right)\right].
\end{equation}
\end{enumerate}

{In this paper, we assume that perfect
instantaneous CSI of the intended receiver is available at the transmitter \cite{Khisti2010TIT,Khisti2010TIT_2,Wu2012TVT,Bashar2011Cletter,Bashar2012TCom,Gursoy2012TCom,Liu2014TSP,Zeng2016TWC}.
This assumption applies for scenarios where
the intended receiver is static or the mobility of the intended receiver is low.
In this case, the coherence time of the intended receiver's
instantaneous channel is large.  Therefore, the instantaneous CSI can be
estimated accurately at the receiver based on training sequences and then
be sent back to the transmitter through dedicated feedback links.
In time division duplex systems, the instantaneous CSI can alternatively be
obtained by exploiting the reciprocity of uplink and downlink.

For the eavesdropper, we consider two cases. The first case is that
the eavesdropper is an idle user of the system and
the transmitter intends to send a private message to a particular user of the system while
regarding the other users as eavesdroppers. In this case, we assume perfect instantaneous
CSI of eavesdropper can be obtained at the transmitter \cite{Khisti2010TIT,Khisti2010TIT_2,Gursoy2012TCom,Liu2014TSP,Zeng2016TWC}.
The second case is that the eavesdropper is always passive and does not transmit.
In this case, we assume that only statistical CSI of the eavesdropper is available
at the transmitter \cite{Shafiee2007ISIT,Li2011TWC,Rezki2014TWC,Wu2016TIT}.}

The goal of this paper is to optimize the transmit precoding matrix $\mathbf{G}$  for maximization of
 the secrecy rate in (\ref{rate}) or (\ref{rate_erg})
when the transmit symbols ${\mathbf{x}}_a $ are drawn from a discrete constellation set with $M$ equiprobable points
such as $M$-ary quadrature amplitude modulation (QAM) and $N_t$ is large.

\section{Low Complexity Precoder Design with Instantaneous CSI of the Eavesdropper}
In this section, we first provide some useful definitions which will be used in the
subsequent analysis. Then, we analyze the rate loss
of the GSVD design \cite{Bashar2012TCom} compared to the maximal rate for finite alphabet inputs in the high SNR regime.
Finally, we propose a PG-GSVD precoder to compensate this performance loss with low complexity.

\subsection{Some Useful Definitions} \label{sec:definition}
Let us introduce some useful definitions for the subsequent analysis.
\begin{definition}
Similar to \cite{Bashar2012TCom,Khisti2010TIT_2}, we define the following subspaces
\begin{eqnarray*}
\mathcal{S}_{ba}    & = {\rm null}\left(\mathbf{H}_{ba}\right)^{\bot}&\cap~\,\,{\rm null}\left(\mathbf{H}_{ea}\right)  \\
\mathcal{S}_{be}  &={\rm null}\left(\mathbf{H}_{ba}\right)^{\bot}&\cap~\,\,{\rm null}\left(\mathbf{H}_{ea}\right)^{\bot}\\
\mathcal{S}_{ea}   &= {\rm null}\left(\mathbf{H}_{ba}\right)&\cap~\,\,{\rm null}\left(\mathbf{H}_{ea}\right)^{\bot}\\
\mathcal{S}_{n}   &={\rm null}\left(\mathbf{H}_{ba}\right)&\cap~\,\,{\rm null}\left(\mathbf{H}_{ea}\right).
\end{eqnarray*}

\end{definition}

\noindent Define $k={\rm rank}\left(\left[\begin{array}{cc}
\mathbf{H}_{ba}^{H} & \mathbf{H}_{ea}^{H}\end{array}\right]^{H}\right)$ and hence ${\rm dim}\left(\mathcal{S}_{n}\right)= N_t -k$.
In addition, define $r={\rm dim}\left(\mathcal{S}_{ba}\right)$ and $s={\rm dim}\left(\mathcal{S}_{be}\right)$.
Therefore, ${\rm dim}\left(\mathcal{S}_{ea}\right)=k-r-s$.

\begin{definition}
Following \cite{Khisti2010TIT_2}, we define  the GSVD of the pair $\left(\mathbf{H}_{ba}, \mathbf{H}_{ea} \right)$
as follows:
\begin{equation} \label{eq:sigma_ba}
\mathbf{H}_{ba}= {\mathbf{U}}_{ba}~\boldsymbol{\Sigma}_{ba}
\kbordermatrix {~ & k & N_t -k \cr
		               ~ &\boldsymbol{\Omega}^{-1}  &  \mathbf{0} \cr}
~{\mathbf{U}}_{a}^{H}
\end{equation}
\begin{equation} \label{eq:sigma_ea}
\mathbf{H}_{ea}= {\mathbf{U}}_{ea}~\boldsymbol{\Sigma}_{ea}
\kbordermatrix {~ & k & N_t -k \cr
		               ~ &\boldsymbol{\Omega}^{-1}  &  \mathbf{0} \cr}
~{\mathbf{U}}_{a}^{H}
\end{equation}
where ${\mathbf{U}}_{a} \in \mathbb{C}^{N_t \times N_t}$, ${\mathbf{U}_{ba}} \in \mathbb{C}^{N_r \times N_r}$,
and ${\mathbf{U}}_{ea}\in \mathbb{C}^{N_e \times N_e}$ are unitary matrices.
$\boldsymbol{\Omega} \in \mathbb{C}^{k\times k}$ is a non-singular matrix with diagonal elements $\omega_i$, $i=1,\ldots,k$.
 $\boldsymbol{\Sigma}_{ba}\in\mathbb{C}^{N_r \times k}$ and $\boldsymbol{\Sigma}_{ea}\in \mathbb{C}^{N_e\times k}$ can be expressed as

\begin{equation}
\boldsymbol{\Sigma}_{ba} = \kbordermatrix {~		   & k-r-s	 & s 		& r 		\cr
										N_r - r - s & \mathbf{0} 	& \mathbf{0}  		& \mathbf{0} 	\cr
										s                 & \mathbf{0}  	& \mathbf{D}_b 	& \mathbf{0} 	\cr
										r                 & \mathbf{0}  	& \mathbf{0}      	& \mathbf{I}_r 	\cr}
\end{equation}

\begin{equation}
\boldsymbol{\Sigma}_{ea} = \kbordermatrix{~		   & k-r-s	 & s 		& r 		\cr
										k - r - s      & \mathbf{I}_{k-r-s} 	& \mathbf{0} 		& \mathbf{0} 	\cr
										s                 & \mathbf{0} 	& \mathbf{D}_e 	& \mathbf{0} 	\cr
										N_e-k+r    & \mathbf{0} 	& \mathbf{0}     	& \mathbf{0}	\cr}
\end{equation}

\end{definition}

\noindent where $\mathbf{D}_b={\rm diag}\left(\left[b_1,\ldots,b_s\right]\right) \in \mathbb{R}^{s \times s} $ and
$\mathbf{D}_e={\rm diag}\left(\left[e_1,\ldots,e_s\right]\right) \in \mathbb{R}^{s \times s}$
are diagonal matrices with real valued entries. The diagonal elements of $\mathbf{D}_b$ and $\mathbf{D}_e$ are ordered
as follows:
\[0<b_1\leq b_2 \leq \ldots \leq b_s < 1\] \[1>e_1\geq e_2 \geq \ldots \geq e_s > 0\] and \[b_p^2 + e_p^2 =1,~\mbox{for}~p=1,\ldots,s .\]

\subsection{Performance Loss of the GSVD Design}
The precoding matrix for the GSVD design can be expressed as \cite{Bashar2012TCom}
\begin{equation}
\mathbf{G} = \mathbf{U}_{a} \mathbf{A} \mathbf P^{\frac{1}{2}}
\label{eq:precoding_matrix}
\end{equation}
where  $\mathbf{P} = {\rm diag} \left(p_1,\ldots,p_{N_t}\right)$ represents
a diagonal power allocation matrix and $\mathbf{A}$ is given by
\begin{equation}
\mathbf{A} = \kbordermatrix{
~ 		& k 							& N_t - k 	\cr
k		& \boldsymbol{\Omega}		& \mathbf{0}		\cr
N_t -k	& \mathbf{0}						& \mathbf{0}		\cr
}.
\end{equation}
For the GSVD precoder design in (\ref{eq:precoding_matrix}), the
received signals ${\mathbf{y}}_b$ and ${\mathbf{y}}_e$ in (\ref{main}) and (\ref{eave})
can be re-expressed as
\begin{eqnarray}
\mathbf{\tilde{y}}_{b} &=&  ~\boldsymbol{\Sigma}_{ba}
\kbordermatrix {~ & k     	&   N_t -k 	\cr
		                ~ & \mathbf{I}_k 	&  \mathbf{0}		\cr}
~\mathbf{P}^{\frac{1}{2}}~\mathbf{x}_a + \mathbf{\tilde{n}}_{b} \label{eq:bob_precoded}\\
\mathbf{\tilde{y}}_{e} &=& ~\boldsymbol{\Sigma}_{ea}
\kbordermatrix {~ & k     	&  N_t -k 	\cr
		                ~ & \mathbf{I}_k 	&   \mathbf{0} 		\cr}
~\mathbf{P}^{\frac{1}{2}}~ \mathbf{x}_a +\mathbf{\tilde{n}}_{e}  \label{eq:eve_precoded}
\end{eqnarray}
where $\tilde{\mathbf y}_{b} = {\mathbf{U}}_{ba}^H \mathbf{y}_b$, $\tilde{\mathbf y}_{e} ={\mathbf{U}}_{ea}^H \mathbf{y}_e$,
 $\tilde{\mathbf{n}}_{b} = {\mathbf{U}}_{ba}^H \mathbf{n}_b$, and $\tilde{\mathbf n}_{e} = {\mathbf{U}}_{ea}^H \mathbf{n}_e$.

Define $N_1 = {\rm rank}\left( {\mathbf{H}}_{ba} \right)$ and $N_2 = {\rm rank}\left( {\mathbf{H}}_{ea} \right)$.
In the following theorem, we analyze the performance of the GSVD  design for finite alphabet inputs in the
high SNR regime.
\begin{theorem}\label{GSVD_loss}
In the high SNR regime  ($P \rightarrow \infty$), for the GSVD design in (\ref{eq:precoding_matrix}),
the achievable secrecy rate $R_{\rm sec, high}$ for finite alphabet signals
is upper bounded by
\begin{equation}\label{rate_high}
R_{\rm sec, high} \leq  N_1 \log_2 M  \ \textrm{b/s/Hz}.
\end{equation}
\begin{proof}
See Appendix \ref{GSVD_loss_proof}.
\end{proof}
\end{theorem}

Theorem \ref{GSVD_loss} indicates that the GSVD design may result in
a severe performance loss for finite alphabet inputs in the high SNR regime.
For example, if $N_t > N_r$, which is a typical scenario for large-scale MIMO systems \cite{Marzetta2010TWC,Larsson2014CM},
the GSVD design will cause a rate loss of at least $(N_t -N_r) \log_2 M$ b/s/Hz compared to
the maximal rate in the high SNR regime. The precoder design in \cite{Wu2012TVT} avoids this performance loss by directly optimizing
the precoder matrix $\mathbf{G}$.  However, this results in an intractable implementation complexity for large-scale MIMO
systems. Inspired by the idea of decoupling and grouping of point-to-point MIMO channels
for finite alphabet inputs \cite{Mohammed2011TIT,Ketseoglou2015TWC,Wu2016ICC}, we propose a PG-GSVD
precoder design that prevents the performance loss of the GSVD design
while retaining a low complexity  in large-scale MIMOME channels.

\subsection{PG-GSVD Precoder Design} \label{sec:PG-GSVD}
As indicated in \cite{Wu2016ICC}, in order to decouple the MIMO channels
into $N_t$ parallel subchannels, the MIMO channel matrix has to be an $N_t \times N_t$ matrix.
However,  $\boldsymbol{\Sigma}_{ba}$ and $\boldsymbol{\Sigma}_{ea}$ in (\ref{eq:bob_precoded}) and (\ref{eq:eve_precoded}) are
$N_r \times N_t$ and $N_e \times N_t$ matrices, respectively. As a result,
we need to add to or remove from $\mathbf{\tilde{y}}_{b}$, $\boldsymbol{\Sigma}_{ba}$,
$\mathbf{\tilde{y}}_{e}$, and $\boldsymbol{\Sigma}_{ea}$ some zeros in (\ref{eq:bob_precoded}) and (\ref{eq:eve_precoded}).
To this end, we define
\begin{align}
\mathbf{\hat{y}}_{b} = \kbordermatrix{
~ & ~ \cr
k - r -s  & \mathbf{0}  \cr
r + s & \tilde{\tilde{\mathbf{y}}}_{b}^H  \cr
 N_t - k & \mathbf{0}},
\end{align}
where $\tilde{\tilde{\mathbf{y}}}_{b} \in \mathbb{C}^{(r + s) \times 1}$
is composed of the last $r + s$ elements of $\mathbf{\tilde{y}}_{b}$.
Furthermore, we define $\boldsymbol{\omega} = \left[ \omega_1,\cdots,\omega_k \ \mathbf{0}^{T} \right]^H \in \mathbb{C}^{N_t \times 1}$,
$\mathbf{\hat{y}}_{e} = \left[\tilde{\mathbf y}_{e}^H \ \mathbf{0}^{T} \right]^H \in \mathbb{C}^{N_t \times 1}$,
$\mathbf{\hat{n}}_{b} \sim {\cal CN}(0, \; \sigma_b^2 {\bf{I}}_{N_t})$, and $\mathbf{\hat{n}}_{e}  \sim {\cal CN}(0, \; \sigma_e^2 {\bf{I}}_{N_t}) $.
Define two diagonal matrices
\begin{equation}\label{eq:sigma_ba_hat}
\boldsymbol{\hat{\Sigma}}_{ba} = \kbordermatrix {~		   &  k -r-s	 & s 		& r 	& N_t - k	\cr
										k - r - s & \mathbf{0} 	& \mathbf{0}  		& \mathbf{0} & \mathbf{0}	\cr
										s                 & \mathbf{0}  	& \mathbf{\hat{D}}_b 	& \mathbf{0} & \mathbf{0}	\cr
										r                 & \mathbf{0}  	& \mathbf{0}      	& \mathbf{R}_r 	& \mathbf{0}  \cr
                                       N_t -k                 & \mathbf{0}  	& \mathbf{0}      	& \mathbf{0} 	& \mathbf{0}  \cr  }
\end{equation}

\begin{equation} \label{eq:sigma_ea_hat}
\boldsymbol{\hat{\Sigma}}_{ea} = \kbordermatrix{~		   & k-r-s	 & s 		& N_t - k + r		\cr
										k - r - s      & \mathbf{R}_{k-r-s} 	& \mathbf{0} 		& \mathbf{0}  	\cr
										s                 & \mathbf{0} 	& \mathbf{\hat{D}}_e 	& \mathbf{0} 	\cr
										N_t -k+r    & \mathbf{0} 	& \mathbf{0}     	& \mathbf{0} 	\cr
                                         }
\end{equation}
where the elements of $\mathbf{\hat{D}}_b$,  $\mathbf{R}_r$,
$\mathbf{R}_{k-r-s}$, and $\mathbf{\hat{D}}_e$
are obtained from the following two equations
\begin{align}
& \left[\boldsymbol{\hat{\Sigma}}_{ba}\right]_{(k-r-s+i)(k-r-s+i)} = \left[\boldsymbol{{\Sigma}}_{ba}\right]_{(N_r -r-s + i)(k-r-s+i)}/\sqrt{\omega_i}, \nonumber \\
 & \hspace{3cm} i = 1,\cdots,s+r  \label{eq:sigma_ba_hat_ele} \\
& \left[\boldsymbol{\hat{\Sigma}}_{ea}\right]_{ii} = \left[\boldsymbol{{\Sigma}}_{ea}\right]_{ii}/ \sqrt{\omega_i}, \  i = 1,\cdots,k-r. \label{eq:sigma_ea_hat_ele}
\end{align}
We divide the transmit signal $\mathbf{x}_a$ into $S$ streams
and let $N_{s}=N_{t}/S$\footnote{For convenience, we assume $N_{s}=N_{t}/S$ is an integer in this paper. If
$N_{t}/S$ is not an integer, we can easily obtain an integer $N_s$ by adding zeros in (\ref{eq:sigma_ba_hat})
and (\ref{eq:sigma_ea_hat}).}.
We define the set $\left\{\ell_{1},\ldots,\ell_{N_{t}}\right\}$
as a permutation of $\left\{1,\ldots,N_{t}\right\}$.
${\bf{P}}_s \in \mathbb{C}^{N_{s} \times N_{s}}$
and $\mathbf{V}_s \in \mathbb{C}^{N_{s} \times N_{s}}$,
$s = 1,\ldots,S$, denote a diagonal
and a unitary matrix, respectively.
$\mathbf{V} \in \mathbb{C}^{N_t \times N_t}$ denotes a unitary matrix.
For the proposed PG-GSVD precoder, we set $\mathbf{G}$ as follows
\begin{equation}
\mathbf{G} = \mathbf{U}_{a} \mathbf{A} \mathbf P^{\frac{1}{2}} \mathbf{V}.
\label{eq:precoding_matrix_gsvd}
\end{equation}

We set
\begin{align}\label{eq:P_pair}
{\left[ {\bf{P}} \right]_{\ell_{j}\ell_{j}} } =  \left[ {{{\bf{P}}}}_s \right]_{ii},
\end{align}
where $i = 1,\ldots,N_{\mathrm s}$, $s=1,\ldots,S$, and $j = (s - 1) N_{\mathrm s} + i$.

Also, we set
\begin{align}\label{eq:V_pair}
&  \left[ {{{\bf{V}}}} \right]_{\ell_i \ell_j} = \nonumber \\
&   \left\{ \begin{array}{l}
   {\left[ {{{\bf{V}}_{s}}} \right]_{mn}} \quad {\rm if} \ i = (s - 1) N_{\mathrm s} + m , \ j = (s - 1) N_{\mathrm s}  + n   \\
0 \qquad \quad\;\; {\rm otherwise}
 \end{array} \right.
\end{align}
where $m = 1,\ldots, N_{s}$, $n = 1,\ldots,N_{s}$, $s = 1,\ldots, S$,
$i =  1,\ldots, N_{t}$, and $j = 1,\ldots, N_{t}$.
Finally, we let
\begin{align}\label{eq:x_pair}
\left[\mathbf{x}_{s}\right]_{i} =  \left[ \mathbf{x}_a \right]_{\ell_{j}} .
\end{align}
Based on (\ref{eq:precoding_matrix_gsvd})--(\ref{eq:x_pair}) and a pairing scheme $\left\{\ell_{1},\ldots,\ell_{N_{t}}\right\}$,
the equivalent received signals
at Bob and Eve can be decoupled as follows
  \begin{eqnarray}
 \left[ \mathbf{\hat{y}}_{b} \right]_{\ell_j} & =  \left[\boldsymbol{\hat\Sigma}_{ba} \right]_{\ell_j\ell_j}  \left[\mathbf{\hat{x}}\right]_{\ell_j}
 + \left[\mathbf{\hat{n}}_{b} \right]_{\ell_j}   \label{eq:bob_precoded_low}   \\
  \left[ \mathbf{\hat{y}}_{e} \right]_{\ell_j} & =  \left[\boldsymbol{\hat\Sigma}_{ea} \right]_{\ell_j\ell_j}  \left[\mathbf{\hat{x}}\right]_{\ell_j}
  + \left[\mathbf{\hat{n}}_{e} \right]_{\ell_j}
   \label{eq:eve_precoded_low}
  \end{eqnarray}
 where
 \begin{align}\label{eq:x_hat}
   \left[\mathbf{\hat{x}}\right]_{\ell_j} = \left[ \mathbf{P}_s^{\frac{1}{2}} \mathbf{V}_s \mathbf{x}_s \right]_i
 \end{align}
for $i = 1,\ldots,N_{s}$, $s = 1,\ldots, S$, and  $j = (s - 1) N_{s} + i$.
From (\ref{eq:bob_precoded_low})
 and (\ref{eq:eve_precoded_low}), we observe that the transmit signal
 has been
divided into $S$ independent
 groups. In each group, the equivalent signal dimension is $N_s \times 1$.
 We further define $ \left[\mathbf{\hat{y}}_b \right]_{\ell_j} = \left[\mathbf{y}_{b,s}\right]_i$ and
 $ \left[\mathbf{\hat{y}}_e \right]_{\ell_j} = \left[\mathbf{y}_{e,s}\right]_i$.

Based on (\ref{eq:bob_precoded_low}) and (\ref{eq:eve_precoded_low}), the secrecy rate in (\ref{rate}) can be expressed as
\begin{equation}\label{eq:I_pair}
  R_{\rm sec}({\mathbf{G}}) = \sum\limits_{s = 1}^S \left( I\left( {{\bf{y}}_{b,s}};\mathbf{x}_{s} \right) - I\left({{\bf{y}}_{e,s}};\mathbf{x}_{s} \right) \right).
\end{equation}

  \begin{alg} \label{Gradient_Pair}
      Maximizing $R_{\rm sec}({\mathbf{G}})$  with respect to $\mathbf{P}_s$ and $\mathbf{V}_s$.

    \vspace*{1.5mm} \hrule \vspace*{1mm}
      \begin{enumerate}

    \itemsep=0pt

   {\item Initialize $\mathbf{P}_s$  and $\mathbf{V}_{{s}}^{(0)}$ for  $s = 1,\ldots,S$ with $\tr( \mathbf{A} {\bf{P}} \mathbf{A}^H) =N_t$.
    Set $N_{\rm iter}$ and $\varepsilon$ as the maximum iteration number and a threshold, respectively. }

    \item  Initialize $R_{\rm sec}({\mathbf{G}})^{(1)}$ based on (\ref{eq:I_pair}).  Set counter $n = 1$.

    \item Update ${\mathbf{P}}_{{s}}^{(n)}$ for $s= 1,\ldots,S$ along
    the gradient decent direction ${\nabla _{{\bf{P}}_{s}}} R({\mathbf{G}})$.

  {  \item Normalize $\tr\left({\mathbf{P}}_{{s}}^{(n)}\right)$ to satisfy $\tr( \mathbf{A} {\bf{P}} \mathbf{A}^H) =N_t$.}

    \item Update $\mathbf{V}_{{s}}^{(n)}$ for $s= 1,\ldots,S$ along the gradient descent direction ${\nabla _{{\bf{V}}_{s}}} R({\mathbf{G}})$.

    \item Compute $R_{\rm sec}({\mathbf{G}})^{(n+1)}$ based on (\ref{eq:I_pair}).
    If $R_{\rm sec}({\mathbf{G}})^{(n+1)}  - R_{\rm sec}({\mathbf{G}})^{(n)} > \varepsilon$
    and $n \leq N_{\rm iter}$,  set $n = n + 1$ and repeat Steps $3$--$5$;

     \item Compute $\mathbf{P}$ and $\mathbf{V}$ based on (\ref{eq:P_pair}) and (\ref{eq:V_pair}).
     Set $\qG = \mathbf{U}_{a} \mathbf{A} \mathbf P^{\frac{1}{2}} \mathbf{V}$.

    \vspace*{1mm} \hrule

     \end{enumerate}

   \end{alg}

    \null
    \par

\noindent The gradients of $I\left( {{\bf{y}}_{b,s}};\mathbf{x}_{s} \right)$ and $I\left({{\bf{y}}_{e,s}};\mathbf{x}_{s} \right)$
with respect to $\mathbf{P}_{s}$ and $\mathbf{V}_{s}$ can be found in
 \cite[Eq. (22)]{Palomar2006TIT}, based on  which an iterative algorithm  can be derived
for maximizing  $ R_{\rm sec}({\mathbf{G}})$, as given in Algorithm 1.

\textit{Remark 1:} For precoder design for the MIMO wiretap channel with finite alphabet inputs,
 the computational complexity is dominated \cite{zeng2012linear} by the computation
 of the mutual information in \cite[Eq. (12)]{Wu2012TVT} and \cite[Eq. (13)]{Wu2012TVT}.
 {We note that the expectations over the noise vector in \cite[Eq. (12)]{Wu2012TVT} and \cite[Eq. (13)]{Wu2012TVT}
 can be evaluated by an accurate approximation as in \cite[Prop. 2]{zeng2012linear}. Therefore, the corresponding computational complexities are negligible.}
Also, we note that computing
 the expectation over $\mathbf{x}_a$ for the mutual information in   \cite[Eq. (12)]{Wu2012TVT} $\,$ and $\,$ \cite[Eq. (13)]{Wu2012TVT}  $\,$ and $\,$ the
 corresponding $\,$ mean $\,$ square $\,$ error $\,$ (MSE) $\,$ matrix  in \cite[Eq. (17)]{Wu2012TVT}  and  \cite[Eq. (18)]{Wu2012TVT} involves
 additions over the modulation signal space
which scales exponentially with the number of transmit antennas.  The computational complexities of other operations, such as
the matrix product and the GSVD decomposition, are polynomial functions of the number of transmit and receive antennas.
Therefore, for ease of analysis, we just consider   the computational complexity of
calculating the mutual information and the MSE matrix for comparing the complexities of Algorithm \ref{Gradient_Pair}
and the complete-search design in \cite{Wu2012TVT}.
For large $N_{t}$, the computational complexity of the complete-search design  in \cite{Wu2012TVT}
is dominated by number of additions needed for calculating the mutual information and
the MSE matrix in  \cite[Eq. (12)]{Wu2012TVT}, \cite[Eq. (13)]{Wu2012TVT},
\cite[Eq. (17)]{Wu2012TVT}, and \cite[Eq. (18)]{Wu2012TVT}.
 Accordingly, the  computational complexity of the complete-search design  scales linearly with
$M^{2 N_{t} }$. In contrast,  the computational complexity of
calculating the mutual information and the MSE matrix in Algorithm \ref{Gradient_Pair}
 based on  (\ref{eq:I_pair}) and  \cite[Eq. (22)]{Palomar2006TIT} grows linearly with $ S M^{2 N_{s} }$.
As a result, the computational complexity of Algorithm  \ref{Gradient_Pair}  can be
significantly lower than that of the complete-search design when the number of transmit
antennas is large.

\textit{Remark 2:} We note that Algorithm \ref{Gradient_Pair} never decreases
the secrecy rate $R_{\rm sec}({\mathbf{G}})$ in any iteration, see Step 6.
Also, for finite alphabet input signals, we know that
the secrecy rate $R_{\rm sec}({\mathbf{G}})$ is upper-bounded.
This indicates that  Algorithm \ref{Gradient_Pair} generates
a non-decreasing sequence which is upper-bounded.
Therefore, Algorithm \ref{Gradient_Pair} is convergent.  Due to the non-convexity of
the objective function $R_{\rm sec}({\mathbf{G}})$,  Algorithm \ref{Gradient_Pair}
 will reach a local optimal point of the secrecy rate in general. As a result,
 we implement Algorithm \ref{Gradient_Pair} for several random
initializations for $\mathbf{P}_s$ and $\mathbf{V}_s$ and choose the result that achieves the highest secrecy
rate as the final design solution \cite{Perez-Cruz2010TIT,Wu2015TWC}.

For the PG-GSVD design in (\ref{eq:precoding_matrix_gsvd}), we have the following theorem.

\begin{theorem}\label{PG-GSVD_high}
If the inequality $ (k - N_2) N_s \geq N_t$ holds,
then we can always find a permutation $\left\{\ell_{1},\ldots,\ell_{N_{t}}\right\}$  for the PG-GSVD
design in (\ref{eq:precoding_matrix_gsvd}), which achieves  $R_{\rm sec, high} =  N_t \log_2 M $ b/s/Hz
in the high SNR regime.
\begin{proof}
See Appendix \ref{Proof_PG-GSVD_high}.
\end{proof}
\end{theorem}

The algorithm in \cite{Wu2012TVT} is equivalent to setting $N_s = N_t$ in Algorithm \ref{Gradient_Pair}. Therefore, as long as $k - N_2 \neq 0$, it can
compensate the performance loss of the GSVD design and achieve the saturation rate  $N_t \log_2 M$ b/s/Hz in the high SNR regime,
as shown in \cite[Figs. 1, 2]{Wu2012TVT}. However, in this case, the computational complexity of the algorithm in \cite{Wu2012TVT} grows
exponentially with $N_t$.  This is prohibitive in large-scale MIMO systems.
For typical large-scale MIMO systems, we have $N_t > N_2$ \cite{Marzetta2010TWC,Larsson2014CM}, which implies $k - N_2 \neq 0$.
As a result, by properly choosing $N_s$, we can reach a favorable trade-off between complexity and secrecy rate
performance\footnote{We note that as long as $k - N_2 \neq 0$, the proposed PG-GSVD design can also be
applied in cases  where the transmitter is equipped with fewer antennas than the receiver.}.

To better illustrate the GSVD design, the precoder design in \cite{Wu2012TVT}, and the PG-GSVD design,
we provide the following example.

\begin{example}
\label{4_3_2} We consider a MIMOME model with $N_t = 4$, $N_r = 3$, $N_e = 2$, and $N_s =2$.
Let ${\bf{x}}_a = [x_1,x_2,x_3,x_4]^T $.  $\left[\mathbf{\hat{D}}_b\right]_{ii}$,
$\left[\mathbf{R}_r\right]_{ii}$, $\left[\mathbf{R}_{k-r-s}\right]_{ii}$, and
$\left[\mathbf{\hat{D}}_e\right]_{ii}$ in (\ref{eq:sigma_ba_hat}) and (\ref{eq:sigma_ea_hat})
are denoted as $D_{b,i}$, $R_{r,i}$, $R_{k - r - s,i}$, and
$D_{e,i}$, respectively. $\left[\mathbf{V}\right]_{ij}$ is denoted as $v_{ij}$.
Based on (\ref{eq:sigma_ba_hat}) and (\ref{eq:sigma_ea_hat}), the equivalent received signals $\mathbf{\hat{y}}_{b}$
and $\mathbf{\hat{y}}_{e}$ for the GSVD design are given by
\begin{subequations}
\begin{equation}
\begin{array}{l}
 {\bf{\hat y}}_b  =    \left[ {\begin{array}{*{20}c}
   0 & 0 & 0 & 0  \\
   0 & {D_{b,1} } & 0 & 0  \\
   0 & 0 & {R_{r,1} } & 0  \\
   0 & 0 & 0 & {R_{r,2} }  \\
\end{array}} \right]\left[ {\begin{array}{*{20}c}
   {p_1 } & 0 & 0 & 0  \\
   0 & {p_2 } & 0 & 0  \\
   0 & 0 & {p_3 } & 0  \\
   0 & 0 & 0 & {p_4 }  \\
\end{array}} \right] \\
\hspace{1cm} \times \left[ \begin{array}{l}
 x_1  \\
 x_2  \\
 x_3  \\
 x_4  \\
 \end{array} \right]   + \widehat{\bf{n}}_b
\end{array}
 \end{equation}
\begin{equation} \label{eq:y_4_exp_b}
\hspace{-3.9cm} = \left[ {\begin{array}{*{20}c}
   0  \\
   {p_2 D_{b,1} x_2 }  \\
   {p_3 R_{r,1} x_3 }  \\
   {p_4 R_{r,2} x_4 }  \\
\end{array}} \right] + \widehat{\bf{n}}_b
\end{equation}
\end{subequations}

\begin{subequations}
\begin{equation}
\begin{array}{l}
{\bf{\hat y}}_e   =  \left[ {\begin{array}{*{20}c}
   {R_{k - r - s,1} } & 0 & 0 & 0  \\
   0 & {D_{e,1} } & 0 & 0  \\
   0 & 0 & 0 & 0  \\
   0 & 0 & 0 & 0  \\
\end{array}} \right]\left[ {\begin{array}{*{20}c}
   {p_1 } & 0 & 0 & 0  \\
   0 & {p_2 } & 0 & 0  \\
   0 & 0 & {p_3 } & 0  \\
   0 & 0 & 0 & {p_4 }  \\
\end{array}} \right] \\
\hspace{1cm} \times \left[ \begin{array}{l}
 x_1  \\
 x_2  \\
 x_3  \\
 x_4  \\
 \end{array} \right]  + \widehat{\bf{n}}_e   \\
 \end{array}
\end{equation}
\begin{equation}  \label{eq:y_4_exp_e}
\hspace{-3.2cm}  = \left[ {\begin{array}{*{20}c}
   {p_1 R_{k - r - s,1} x_1 }  \\
   {p_2 D_{e,1} x_2 }  \\
   0  \\
   0  \\
\end{array}} \right]  + \widehat{\bf{n}}_e.
\end{equation}
\end{subequations}

We observe from (\ref{eq:y_4_exp_b}) and (\ref{eq:y_4_exp_e}) that the GSVD design decouples
the original MIMO wiretap channel into four parallel subchannels.
In subchannel 3 and 4, $x_3$ and $x_4$ are received by Bob but not by Eve, respectively.
In subchannel 1,
$x_1$ is received by Eve but not by Bob. In subchannel 2, according to \cite[Eq. (12)]{Bashar2012TCom}, if $D_{b,1} < D_{e,1}$,
we set $p_2 = 0$.  As a result, for the GSVD design,
$x_1$ is definitely not received  by Bob and $x_2$ may also not be received by Bob.

The equivalent received signals $\mathbf{\hat{y}}_{b}$
and $\mathbf{\hat{y}}_{e}$ for the design in \cite{Wu2012TVT} are given by
\begin{subequations}
\begin{equation}
\begin{array}{l}
 {\bf{\hat y}}_b  =   \left[ {\begin{array}{*{20}c}
   0 & 0 & 0 & 0  \\
   0 & {D_{b,1} } & 0 & 0  \\
   0 & 0 & {R_{r,1} } & 0  \\
   0 & 0 & 0 & {R_{r,2} }  \\
\end{array}} \right]\left[ {\begin{array}{*{20}c}
   {p_1 } & 0 & 0 & 0  \\
   0 & {p_2 } & 0 & 0  \\
   0 & 0 & {p_3 } & 0  \\
   0 & 0 & 0 & {p_4 }  \\
\end{array}} \right]  \\
\hspace{1cm} \times \left[ {\begin{array}{*{20}c}
   {v_{11} } & {v_{12} } & {v_{13} } & {v_{14} }  \\
   {v_{21} } & {v_{22} } & {v_{23} } & {v_{24} }  \\
   {v_{31} } & {v_{32} } & {v_{33} } & {v_{34} }  \\
   {v_{41} } & {v_{42} } & {v_{43} } & {v_{44} }  \\
\end{array}} \right]\left[ \begin{array}{l}
 x_1  \\
 x_2  \\
 x_3  \\
 x_4  \\
 \end{array} \right] + \widehat{\bf{n}}_b
 \end{array}
 \end{equation}
\begin{equation} \label{eq:y_4_exp_b_G}
\hspace{-2.9cm}  = \left[ {\begin{array}{*{20}c}
   0  \\
   {p_2 D_{b,1} \sum\limits_{i = 1}^4 {v_{2i} x_i } }  \\
   {p_3 R_{r,1} \sum\limits_{i = 1}^4 {v_{3i} x_i } }  \\
   {p_4 R_{r,2} \sum\limits_{i = 1}^4 {v_{4i} x_i } }  \\
\end{array}} \right] + \widehat{\bf{n}}_b
 \end{equation}
 \end{subequations}

\begin{subequations}
\begin{equation}
\begin{array}{l}
{\bf{\hat y}}_e  =   \left[ {\begin{array}{*{20}c}
   {R_{k - r - s,1} } & 0 & 0 & 0  \\
   0 & {D_{e,1} } & 0 & 0  \\
   0 & 0 & 0 & 0  \\
   0 & 0 & 0 & 0  \\
\end{array}} \right]\left[ {\begin{array}{*{20}c}
   {p_1 } & 0 & 0 & 0  \\
   0 & {p_2 } & 0 & 0  \\
   0 & 0 & {p_3 } & 0  \\
   0 & 0 & 0 & {p_4 }  \\
\end{array}} \right] \\
\hspace{1cm} \times \left[ {\begin{array}{*{20}c}
   {v_{11} } & {v_{12} } & {v_{13} } & {v_{14} }  \\
   {v_{21} } & {v_{22} } & {v_{23} } & {v_{24} }  \\
   {v_{31} } & {v_{32} } & {v_{33} } & {v_{34} }  \\
   {v_{41} } & {v_{42} } & {v_{43} } & {v_{44} }  \\
\end{array}} \right]\left[ \begin{array}{l}
 x_1  \\
 x_2  \\
 x_3  \\
x_4  \\
 \end{array} \right] +  \widehat{\bf{n}}_e
 \end{array}
 \end{equation}
 \begin{equation}\label{eq:y_4_exp_e_G}
\hspace{-2.1cm} = \left[ {\begin{array}{*{20}c}
   {p_1 R_{k - r - s,1} \sum\limits_{i = 1}^4 {v_{1i} x_i } }  \\
   {p_2 D_{e,1} \sum\limits_{i = 1}^4 {v_{2i} x_i } }  \\
   0  \\
   0  \\
\end{array}} \right] +  \widehat{\bf{n}}_e
 \end{equation}
  \end{subequations}

We observe from (\ref{eq:y_4_exp_b_G}) and (\ref{eq:y_4_exp_e_G})
that for the design in \cite{Wu2012TVT}, by setting $p_1 = p_2 =0$,
$x_1$, $x_2$, $x_3$, and $x_4$ are combined and transmitted
along subchannel 3 and subchannel 4, which can be received by Bob but
not by Eve. However, in this case, we need to calculate the expectation
over $\left(x_1,x_2,x_3,x_4\right)$ for evaluating the mutual information
and the MSE matrix, which requires $4^{2 \times 4} = 65536$ additions for quadrature phase shift keying (QPSK)
inputs.

The equivalent received signals $\mathbf{\hat{y}}_{b}$
and $\mathbf{\hat{y}}_{e}$ for the PG-GSVD design are given by
\begin{subequations}
\begin{equation}
\begin{array}{l}
{\bf{\hat y}}_b  =
 \left[ {\begin{array}{*{20}c}
   0 & 0 & 0 & 0  \\
   0 & {D_{b,1} } & 0 & 0  \\
   0 & 0 & {R_{r,1} } & 0  \\
   0 & 0 & 0 & {R_{r,2} }  \\
\end{array}} \right]\left[ {\begin{array}{*{20}c}
   {p_1 } & 0 & 0 & 0  \\
   0 & {p_2 } & 0 & 0  \\
   0 & 0 & {p_3 } & 0  \\
   0 & 0 & 0 & {p_4 }  \\
\end{array}} \right] \\
\hspace{1cm} \times    \left[ {\begin{array}{*{20}c}
   {v_{11} } & 0 & 0 & {v_{14} }  \\
   0 & {v_{22} } & {v_{23} } & 0  \\
   0 & {v_{32} } & {v_{33} } & 0  \\
   {v_{41} } & 0 & 0 & {v_{44} }  \\
\end{array}} \right]\left[ \begin{array}{l}
 x_1  \\
 x_2  \\
 x_3  \\
 x_4  \\
 \end{array} \right] + \widehat{\bf{n}}_b
  \end{array}
 \end{equation}
 \begin{equation}\label{eq:y_4_exp_b_pro}
\hspace{-1.8cm} = \left[ {\begin{array}{*{20}c}
   0  \\
   {p_2 D_{b,1} \left( {v_{22} x_2  + v_{23} x_3 } \right)}  \\
   {p_3 R_{r,1} \left( {v_{32} x_2  + v_{33} x_3 } \right)}  \\
   {p_4 R_{r,2} \left( {v_{41} x_1  + v_{44} x_4 } \right)}  \\
\end{array}} \right] +   \widehat{\bf{n}}_b
 \end{equation}
  \end{subequations}

\begin{subequations}
\begin{equation}
\begin{array}{l}
 {\bf{\hat y}}_e  = \left[ {\begin{array}{*{20}c}
   {R_{k - r - s,1} } & 0 & 0 & 0  \\
   0 & {D_{e,1} } & 0 & 0  \\
   0 & 0 & 0 & 0  \\
   0 & 0 & 0 & 0  \\
\end{array}} \right]\left[ {\begin{array}{*{20}c}
   {p_1 } & 0 & 0 & 0  \\
   0 & {p_2 } & 0 & 0  \\
   0 & 0 & {p_3 } & 0  \\
   0 & 0 & 0 & {p_4 }  \\
\end{array}} \right] \\
\hspace{1cm} \times \left[ {\begin{array}{*{20}c}
   {v_{11} } & 0 & 0 & {v_{14} }  \\
   0 & {v_{22} } & {v_{23} } & 0  \\
   0 & {v_{32} } & {v_{33} } & 0  \\
   {v_{41} } & 0 & 0 & {v_{44} }  \\
\end{array}} \right]\left[ \begin{array}{l}
 x_1  \\
 x_2  \\
 x_3  \\
 x_4  \\
 \end{array} \right] + \widehat{\bf{n}}_e
   \end{array}
 \end{equation}
 \begin{equation}\label{eq:y_4_exp_e_pro}
 \hspace{-1.2cm}  = \left[ {\begin{array}{*{20}c}
   {p_1 R_{k - r - s,1} \left( {v_{11} x_1  + v_{14} x_4 } \right)}  \\
   {p_2 D_{e,1} \left( {v_{22} x_2  + v_{23} x_3 } \right)}  \\
   0  \\
   0  \\
\end{array}} \right] + \widehat{\bf{n}}_e
 \end{equation}
  \end{subequations}

We observe from (\ref{eq:y_4_exp_b_pro}) and (\ref{eq:y_4_exp_e_pro}) that for the
PG-GSVD design, $x_1$ and $x_4$ are combined and transmitted over subchannel 4.
For $x_2$, even if $D_{b,1} < D_{e,1}$, it can be combined with $x_3$ and transmitted over  subchannel 3.
For the PG-GSVD design, by setting $p_1 = p_2 =0$, $x_1$, $x_2$, $x_3$, and $x_4$  can  also be received by Bob but
not by Eve. As a result, the PG-GSVD design compensates
the performance loss caused by the GSVD design.  In this case,  we only need to calculate the expectation
over $\left(x_1, x_4\right)$ and $\left(x_2,x_3\right)$ for evaluating the mutual information
and the MSE matrix, which requires $2\times 4^{2 \times 2} = 512$ additions for QPSK inputs.

\end{example}

\section{Low Complexity Precoder Design with Statistical CSI of Eve}
If only statistical CSI of Eve is available at the transmitter,
the PG-GSVD design in Section \ref{sec:PG-GSVD} can not be directly applied
to maximize the ergodic secrecy rate in (\ref{rate_erg}). This is because
in (\ref{rate_erg}), the expectation over all possible realizations
of Eve's channel is needed. It is impossible to find a single precoder matrix to decouple
all channel realizations simultaneously. However, by exploiting the asymptotic approximation
results in \cite{Wu2015TWC}, we can establish the deterministic equivalent channel
of Eve, based on which the PG-GSVD design in Section \ref{sec:PG-GSVD} can be applied directly.
However, this design has two major drawbacks. First, the iterative algorithm
formulated based on this equivalent channel requires the calculation of the
corresponding asymptotic parameters in each iteration (see Step 4 in Algorithm 1 in \cite{Wu2015TWC}),
which increases the computational burden. More importantly, this equivalent channel is obtained
under the assumption that $N_t$ and $N_e$ approach
infinity simultaneously. However, whether this secrecy rate is achievable when
the antenna dimensions are finite is unknown. In this section,
we first propose a low complexity precoder design to maximize an achievable secrecy rate with statistical CSI of Eve.
Then, we discuss the role of AN.
\subsection{PG-GSVD Design with Statistical CSI} \label{sec:PG-GSVD_sta}
In this subsection, we derive a lower bound on (\ref{rate_erg}), which is an achievable secrecy rate
that does not depend on the instantaneous eavesdropper channel but only depends on
the correlation matrices ${\mathbf{R}}_{N_e }$ and $\mathbf{R}_{N_t}$.
Then, we propose a precoder structure  that decomposes these matrices into small dimensions,
which forms the basis for a low complexity precoder design.

Let ${\mathbf{R}}_{N_t } = \mathbf{T}^H \mathbf{T}$. Define the GSVD of the pair $\left(\mathbf{H}_{ba}, \mathbf{T} \right)$
as follows:
\begin{equation} \label{eq:sigma_ba_erg}
\mathbf{H}_{ba}= {\mathbf{{U}}}_{ba, \rm{erg}}~\boldsymbol{\Sigma}_{ba, \rm{erg}}
\kbordermatrix {~ & k_{\rm{erg}} & N_t -k_{\rm{erg}} \cr
		               ~ &{\boldsymbol\Omega^{-1}_{\rm{erg}}}  &  \mathbf{0} \cr}
~{\mathbf{U}}_{a,\rm{erg}}^{H}
\end{equation}
\begin{align} \label{eq:sigma_ea_erg}
 \mathbf{T} = {\mathbf{U}}_{ea, \rm{erg}}~\boldsymbol{\Sigma}_{ea, \rm{erg}}
\kbordermatrix {~ & k_{\rm{erg}} & N_t -k_{\rm{erg}} \cr
		               ~ &{\boldsymbol\Omega^{-1}_{\rm{erg}}}  &  \mathbf{0} \cr}
~{\mathbf{U}}_{a, \rm{erg}}^{H}
\end{align}
where ${\mathbf{{U}}}_{ba, \rm{erg}}$, $\boldsymbol{\Sigma}_{ba, \rm{erg}}$,
${\mathbf{U}}_{a,\rm{erg}}$, ${\mathbf{U}}_{ea, \rm{erg}}$, ${\boldsymbol\Sigma}_{ea, \rm{erg}}$, $\mathbf{A}_{\rm{erg}}$,
$\boldsymbol\Omega_{\rm{erg}}$, $k_{\rm{erg}}$, $r_{\rm{erg}}$, and $s_{\rm{erg}}$ are obtained
by replacing $\mathbf{H}_{ea}$ with $\mathbf{T}$ in Section \ref{sec:definition}.

Also, we define
\begin{align}
& \boldsymbol{{\hat{\Sigma}}}_{ea,\rm{erg}}  = \nonumber  \\
& \kbordermatrix{~		   & k_{\rm erg} -r_{\rm erg} -s_{\rm erg}	 & s_{\rm erg} 		& N_t - k_{\rm erg} + r_{\rm erg}		\cr
										k_{\rm erg} - r_{\rm erg} - s_{\rm erg}      & {\mathbf R}_{\rm erg} 	& \mathbf{0} 		& \mathbf{0}  	\cr
										s_{\rm erg}                 & \mathbf{0} 	& \mathbf{{\hat{D}}}_{e,\rm{erg}} 	& \mathbf{0} 	\cr
										N_t -k_{\rm erg} +r_{\rm erg}    & \mathbf{0} 	& \mathbf{0}     	& \mathbf{0} 	\cr
                                         }
\end{align}
where the elements of ${\mathbf R}_{\rm erg}$ and $\mathbf{{\hat{D}}}_{e,\rm{erg}}$
are obtained from the following equation:
\begin{align}
 \left[\boldsymbol{\hat{\Sigma}}_{ea, \rm{erg}}\right]_{ii} = \frac{\left[\boldsymbol{{{\Sigma}}}_{ea,\rm{erg}}\right]_{ii}}{ \sqrt{\omega_{i,\rm{erg}}}},
\quad i = 1,\ldots,k_{\rm{erg}}-r_{\rm{erg}}
\end{align}
where $\omega_{i,\rm{erg}}$ is the $i$th diagonal element of $\boldsymbol\Omega_{\rm{erg}}$.
Let ${\mathbf{x}}_{q}$ denote the $q$th element of the transmit signal constellation set
and $\mathbf{b}_{pq} = {\mathbf{x}}_{p} - {\mathbf{x}}_{q}$, $p = 1,\ldots,M^{N_t}$, $q = 1,\ldots,M^{N_t}$.
We further define $\left[ \boldsymbol{{\hat{\Sigma}}}_{ea,\rm{erg}} \right]_{\ell_j \ell_j} =  {\left[ {{\boldsymbol{{\hat{\Sigma}}}_{s}}} \right]_{ii}}$
and $ \left[\mathbf{b}_{pq}\right]_{\ell_j \ell_j} = \left[ \mathbf{b}_{s,pq} \right]_{ii}$
for $i = 1,\ldots,N_{s}$, $s = 1,\ldots, S$, and  $j = (s - 1) N_{s} + i$.
$\mathbf P_{\rm{erg}} \in \mathbb{C}^{N_{t} \times N_{t}} $ and $\mathbf{V}_{\rm{erg}} \in \mathbb{C}^{N_{t} \times N_{t}}$
denote a diagonal and a unitary matrix, respectively.
Replace $\mathbf{P}_s$, $\mathbf{V}_s$, $\mathbf{P}$, and $\mathbf{V}$
in (\ref{eq:P_pair}) and (\ref{eq:V_pair}) with
$\mathbf{P}_{s,\rm{erg}} \in \mathbb{C}^{N_{s} \times N_{s}}$,
$\mathbf{V}_{s,\rm{erg}} \in \mathbb{C}^{N_{s} \times N_{s}}$,
$\mathbf P_{\rm{erg}}$, and $\mathbf{V}_{\rm{erg}}$, respectively.
Then, we have the following theorem.
\begin{theorem}\label{PG-GSVD_erg}
By setting $\mathbf{G}_{\rm{erg}} = \mathbf{U}_{a,\rm{erg}} \mathbf{A}_{\rm{erg}} \mathbf P_{\rm{erg}}^{\frac{1}{2}} \mathbf{V}_{\rm{erg}}$,
a lower bound on $\bar{R}_{\rm{sec} }({\mathbf{G}})$ in (\ref{rate_erg}) is given by
\begin{equation}\label{sec_rate_erg_lower_The}
\bar{R}_{\rm{sec} }({\mathbf{G}})\geq \bar{R}_{\rm{sec,\,l}}({\mathbf{G}}) =\sum\limits_{s = 1}^S I\left( {{\bf{y}}_{b,s,\rm{erg}}};\mathbf{x}_{s,\rm{erg}} \right)
 - R_{\rm{eve,\,u}},
\end{equation}
where ${{\bf{y}}_{b,s,\rm{erg}}}$ and $\mathbf{x}_{s,\rm{erg}}$ are obtained  by
replacing $\mathbf{H}_{ea}$ with $\mathbf{T}$ in Section \ref{sec:PG-GSVD}
and $R_{\rm{eve,\,u}}$ is given by
\begin{align}
& R_{\rm{eve,\,u}} = N_t \log M - \frac{1}{M^{N_t} } \sum\limits_{s = 1}^{S} \sum\limits_{p_s = 1}^{M^{N_s}}  \log  \sum\limits_{q_s = 1}^{M^{N_s}} \exp\left( - \frac{\tr({\mathbf{R}}_{N_e}) }{\sigma_e^2} \right. \nonumber \\
& \left. \times  \mathbf{b}_{s,p_s q_s}^H
 \mathbf{V}_{s,\rm{erg}}^H \mathbf P_{s,\rm{erg}}^{\frac{1}{2}}
\boldsymbol{\hat{\Sigma}}_{s}^2 \mathbf P_{s,\rm{erg}}^{\frac{1}{2}} \mathbf{V}_{s,\rm{erg}} \mathbf{b}_{s,p_s q_s} \right). \label{rate_erg_lower_The}
\end{align}
\begin{proof}
See Appendix \ref{Proof_PG-GSVD_erg}.
\end{proof}
\end{theorem}

In Theorem \ref{PG-GSVD_erg}, we use an upper bound on the eavesdropper's rate, and consequently, obtain  a
lower bound on the ergodic secrecy rate. Therefore, maximizing the lower bound on the
ergodic secrecy rate in (\ref{sec_rate_erg_lower_The}) will yield a lower bound on
the achievable ergodic secrecy rate, $\bar{C}_{\rm{sec} }$, in (\ref{secrecy_erg}).

Based on \cite[Eq. (31)]{Xiao2011TSP}, the gradient of $R_{\rm{eve,\,u}}$ with respect to $\mathbf{P}_{s, \rm{erg}}$ is
given by
\begin{align}\label{eq:gradient_p}
\nabla_{\mathbf{P}_{s, \rm{erg}}} R_{\rm{eve,\,u}} = \frac{\log e}{M^{N_s} \sigma_e^2} {\rm{Diag}}
 \left( \sum\limits_{p_s = 1}^{M^{N_s}} \frac{g_{p_sq_s} \mathbf{L}_{p_sq_s}^{T}}{\sum\limits_{q_s = 1}^{M^{N_s}}g_{p_sq_s}} \right)
\end{align}
where
\begin{align}
g_{p_sq_s} = \exp \left( -  \frac{\tr\left({\mathbf{R}_{N_e}}\right)\tr\left(\mathbf{L}_{p_sq_s}\mathbf{P}_{s, \rm{erg}} \right)}{\sigma_e^2} \right)
\end{align}
\begin{align}
\mathbf{L}_{p_sq_s} = \boldsymbol{\hat{\Sigma}}_{s}^2 \mathbf{V}_{s,\rm{erg}} \mathbf{b}_{s,p_sq_s}  \mathbf{b}_{s,p_sq_s}^H \mathbf{V}_{s,\rm{erg}}^H.
\end{align}

The gradient of $R_{\rm{eve,\,u}}$ with respect to $\mathbf{V}_{s, \rm{erg}}$ can be calculated as
\begin{subequations}
\begin{equation}
\begin{array}{l}
 \nabla_{\mathbf{V}_{s, \rm{erg}}} R_{\rm{eve,\,u}} = \frac{\tr\left({\mathbf{R}_{N_e}}\right)\log e}{M^{N_s} \sigma_e^2}
\sum\limits_{p_s = 1}^{M^{N_s}}  \frac{1}{{\sum\limits_{q_s = 1}^{M^{N_s}}g_{p_sq_s}}}    \sum\limits_{q_s = 1}^{M^{N_s}} g_{p_sq_s} \\
\times \nabla_{\mathbf{V}_{s, \rm{erg}}}   \mathbf{b}_{s,p_s q_s}^H
 \mathbf{V}_{s,\rm{erg}}^H \mathbf P_{s,\rm{erg}}^{\frac{1}{2}}
\boldsymbol{\hat{\Sigma}}_{s}^2 \mathbf P_{s,\rm{erg}}^{\frac{1}{2}} \mathbf{V}_{s,\rm{erg}} \mathbf{b}_{s,p_s q_s}
\end{array}
\end{equation}
\begin{equation}
\begin{array}{l}
 \hspace{2.1cm} =  \frac{\tr\left({\mathbf{R}_{N_e}}\right)\log e}{M^{N_s} \sigma_e^2} \\
 \times \sum\limits_{p_s = 1}^{M^{N_s}}\frac{\sum\limits_{q_s = 1}^{M^{N_s}} g_{p_sq_s} \mathbf P_{s,\rm{erg}}^{\frac{1}{2}}
\boldsymbol{\hat{\Sigma}}_{s}^2 \mathbf P_{s,\rm{erg}}^{\frac{1}{2}}
\mathbf{V}_{s,\rm{erg}}  \mathbf{b}_{s,p_s q_s}  \mathbf{b}_{s,p_s q_s}^H }{{\sum\limits_{q_s = 1}^{M^{N_s}}g_{p_sq_s}}}.
\end{array}
\end{equation}
  \end{subequations}

Then, the gradients of $R_{\rm{sec,\,l}}({\mathbf{G}})$ with respect to $\mathbf{P}_{s, \rm{erg}}$
and $\mathbf{V}_{s, \rm{erg}}$ are given by
\begin{align}\label{eq:gradient_RG_p}
\nabla_{\mathbf{P}_{s, \rm{erg}}} \bar{R}_{\rm{sec,\,l}}({\mathbf{G}}) \! = \! \nabla_{\mathbf{P}_{s, \rm{erg}}} I\left( {{\bf{y}}_{b,s,\rm{erg}}};\mathbf{x}_{s,\rm{erg}} \right)
\! - \!  \nabla_{\mathbf{P}_{s, \rm{erg}}} R_{\rm{eve,\,u}}
\end{align}
\begin{align}\label{eq:gradient_RG_V}
\nabla_{\mathbf{V}_{s, \rm{erg}}}  \bar{R}_{\rm{sec,\,l}}({\mathbf{G}}) \! = \! \nabla_{\mathbf{V}_{s, \rm{erg}}} I\left( {{\bf{y}}_{b,s,\rm{erg}}};\mathbf{x}_{s,\rm{erg}} \right)
\! -\!  \nabla_{\mathbf{V}_{s, \rm{erg}}} R_{\rm{eve,\,u}}
\end{align}
where $\nabla_{\mathbf{P}_{s, \rm{erg}}} I\left( {{\bf{y}}_{b,s,\rm{erg}}};\mathbf{x}_{s,\rm{erg}} \right)$
and $\nabla_{\mathbf{V}_{s, \rm{erg}}} I\left( {{\bf{y}}_{b,s,\rm{erg}}};\mathbf{x}_{s,\rm{erg}} \right)$
are given in \cite[Eq. (20)]{Wu2015TWC} and \cite[Eq. (21)]{Wu2015TWC}, respectively.

We propose Algorithm \ref{Gradient_Pair_erg} to maximize the achievable ergodic secrecy rate $\bar{R}_{\rm{sec,\,l}}({\mathbf{G}})$.
It is important to note that the number of
additions required for calculating $\bar{R}_{\rm{sec,\,l}}({\mathbf{G}})$ in (\ref{sec_rate_erg_lower_The}),
$\nabla_{\mathbf{P}_{s, \rm{erg}}} \bar{R}_{\rm{sec,\,l}}({\mathbf{G}})$ in (\ref{eq:gradient_RG_p}), and
$\nabla_{\mathbf{V}_{s, \rm{erg}}}  \bar{R}_{\rm{sec,\,l}}({\mathbf{G}})$ in (\ref{eq:gradient_RG_V})
scales linearly with $ S M^{2 N_{s} }$. Therefore, for large $N_t$, the computational complexity
of Algorithm \ref{Gradient_Pair_erg} scales linearly with $ S M^{2 N_{s} }$.
In contrast, the computational complexity of the secure transmission designs
with statistical CSI in \cite{Bashar2011Cletter,Wu2012TVT} scales linearly with $M^{2 N_{t} }$.

Define $N_3 ={\rm rank}\left( \mathbf{T}\right)$. Based on (\ref{rate_erg_lower_The}), we obtain
the following theorem in the high SNR regime.
\begin{theorem}\label{PG-GSVD_erg_high}
If inequality $ (k_{\rm erg} - N_3) N_s \geq N_t$ holds,
then we can always find a permutation $\left\{\ell_{1},\ldots,\ell_{N_{t}}\right\}$  for the PG-GSVD
design in Theorem \ref{PG-GSVD_erg}, which achieves  $R_{\rm{sec},l,\rm{high}} =  N_t \log_2 M  \ b/s/Hz $
in the high SNR regime, i.e., the maximal secrecy rate is achieved.
\begin{proof}
The proof follows similar steps as those in Appendix B if $\mathbf{H}_{ea}$ is replaced by $\mathbf{T}$.
\end{proof}
\end{theorem}

\textit{Remark 3:} For small dimensional  MIMO channels,  $k_{\rm erg} =N_3 $ may hold.
Thus, it is difficult to find a precoder design that
achieves the maximal asymptotic secrecy rate when only

\begin{alg} \label{Gradient_Pair_erg}
      Maximizing $\bar{R}_{\rm{sec,\,l}}({\mathbf{G}})$  with respect to $\mathbf P_{s,\rm{erg}}$ and $\mathbf{V}_{s,\rm{erg}}$.

    \vspace*{1.5mm} \hrule \vspace*{1mm}
      \begin{enumerate}

    \itemsep=0pt

{    \item Initialize $\mathbf P_{s,\rm{erg}}$  and
    $\mathbf{V}_{s,\rm{erg}}^{(0)}$ for  $s = 1,\ldots,S$ with $\tr(\mathbf{A}_{\rm{erg}}\mathbf{P}_{\rm{erg}} \mathbf{A}_{\rm{erg}}^H) = N_t$.
    Set $N_{\rm iter}$ and $\varepsilon$ as the  maximum number of iterations and a threshold, respectively.}

    \item  Initialize $\bar{R}_{\rm{sec,\,l}}({\mathbf{G}})$  based on (\ref{sec_rate_erg_lower_The}) .
     Set counter $n = 1$.

    \item Update $\mathbf P_{s,\rm{erg}}^{(n)}$ for $s= 1,\ldots,S$ along
    the gradient decent direction $\nabla_{\mathbf{P}_{s, \rm{erg}}} \bar{R}_{\rm{sec,\,l}}({\mathbf{G}})$ based on (\ref{eq:gradient_RG_p}).

{   \item Normalize $\mathbf P_{s,\rm{erg}}^{(n)}$ to satisfy $\tr(\mathbf{A}_{\rm{erg}}\mathbf{P}_{\rm{erg}} \mathbf{A}_{\rm{erg}}^H) = N_t$.}

    \item Update $\mathbf{V}_{s,\rm{erg}}^{(n)}$ for $s= 1,\ldots,S$ along the gradient descent direction
    $\nabla_{\mathbf{V}_{s, \rm{erg}}}  \bar{R}_{\rm{sec,\,l}}({\mathbf{G}})$ based on (\ref{eq:gradient_RG_V}).

    \item Compute $\bar{R}_{\rm{sec,\,l}}({\mathbf{G}})^{(n+1)}$ based on (\ref{sec_rate_erg_lower_The}).
    If $\bar{R}_{\rm{sec,\,l}}({\mathbf{G}})^{(n+1)}  - \bar{R}_{\rm{sec,\,l}}({\mathbf{G}})^{(n)} > \varepsilon$
    and $n \leq N_{\rm iter}$,  set $n = n + 1$ and repeat Steps $3$--$5$;

     \item Compute $\mathbf{P}_{\rm erg}$ and $\mathbf{V}_{\rm erg}$ based on $\mathbf P_{s,\rm{erg}}^{(n)}$ and $\mathbf{V}_{s,\rm{erg}}^{(n)}$.
     Set $\qG_{\rm erg} =\mathbf{U}_{a,\rm{erg}} \mathbf{A}_{\rm{erg}} \mathbf P_{\rm{erg}}^{\frac{1}{2}} \mathbf{V}_{\rm{erg}}$.

    \vspace*{1mm} \hrule

     \end{enumerate}

   \end{alg}

    \null
    \par

\noindent statistical CSI of Eve is available at the transmitter.
In this case,  injection of AN may improve the secrecy rate
in the high SNR regime, as shown in \cite{Bashar2011Cletter,Wu2012TVT}.
On the other hand, for large-scale MIMO channels, it is known that typically $ N_t > N_3 $ holds \cite{Adhikary2013TIT,Yin2014JSTSP,Wu2016TIT}.
Therefore, we have $k_{\rm erg} - N_3 \neq 0 $.
By exploiting this property, we can formulate a PG-GSVD design similar to that
in (\ref{eq:P_high}) and (\ref{eq:pair_high}) to achieve
 the maximal secrecy rate in the high SNR regime by selecting a proper value for $N_s$.
 In this case, AN generation is not necessary in the high regime SNR.

\subsection{AN Generation}
As explained in Remark 3, for $ (k_{\rm erg} - N_3) N_s < N_t$,  AN generation may be
beneficial to  increase the secrecy rate when the transmitter has only statistical CSI of Eve's channel.
In particular, with perfect instantaneous CSI of Bob, the transmitter may construct the AN
along the null space of $\mathbf{H}_{ba}$ as follows \cite{Bashar2011Cletter,Wu2012TVT}:
\begin{equation}\label{eq:precoder_erg}
\mathbf{x}  = \mathbf{G}_{\rm{erg,a}} \mathbf{x}_a + \frac{\sqrt{{P}_{\rm AN}}}{N_t - N_r} \mathbf{V}_b \mathbf{u}
\end{equation}
where $\mathbf{G}_{\rm{erg,a}} \in \mathbb{C}^{N_{t} \times N_{t}}$ is the precoder for the useful
signal, $\mathbf{V}_b  \in \mathbb{C}^{N_{t} \times (N_{t} - N_r)}$ is the null space of $\mathbf{H}_{ba}$,
and $\mathbf{u} \sim {\cal CN}(0, \; {\bf{I}}_{N_t - N_r})$ is the AN.
After obtaining $\mathbf{G}_{\rm{erg,a}}$ via Algorithm \ref{Gradient_Pair_erg},
the AN power ${{P}_{\rm AN}}$ can
be calculated as\footnote{We note that although this power allocation policy and transmitting
the AN in the null space of $\mathbf{H}_{ea}$ are not optimal in general, simulations in
\cite{Bashar2011Cletter,Wu2012TVT} show that such a design  performs well when  the transmitter has perfect statistical CSI of Eve's channel. A more general joint design of the precoder and the AN is provided in \cite{Aghdam2016}. However, such a joint design has a significantly higher
computational complexity when $N_t$ is large. Therefore, for implementation simplicity, in this paper,
we design the precoder and the AN separately as in \cite{Bashar2011Cletter,Wu2012TVT}.}
${{P}_{\rm AN}} = P - {\tr}\left\{ { {\mathbf{G}_{\rm{erg,a}} \mathbf{G}^H_{\rm{erg,a}}}} \right\}$ \cite{Bashar2011Cletter,Wu2012TVT}.

If Alice transmits AN,  Eve's received signal is impaired by
a zero-mean colored Gaussian noise vector with covariance matrix ${{P}_{\rm AN}}/(N_{t} - N_r) \mathbf{H}_{ea} \mathbf{V}_b \mathbf{V}_b^H \mathbf{H}_{ea} ^H
 +  \sigma_e^2 \mathbf{I}_{N_e}$. For large-scale MIMO channels, as $N_t \rightarrow \infty$, we have

\begin{subequations}
\begin{equation} \label{eq:cov_jamming}
\begin{array}{l}
 \mathbf{e}_i^H \mathbf{H}_{ea} \mathbf{V}_b \mathbf{V}_b^H \mathbf{H}_{ea} ^H \mathbf{e}_j  = \\
\hspace{1cm}   \mathbf{e}_i^H {\mathbf{R}}_{N_e }^{\frac{1}{2}} {\mathbf{H}}_w {\mathbf{R}}_{N_t }^{\frac{1}{2}}
\mathbf{V}_b \mathbf{V}_b^H  {\mathbf{R}}_{N_t }^{\frac{1}{2}}{\mathbf{H}}_w^H {\mathbf{R}}_{N_e}^{\frac{1}{2}} \mathbf{e}_j
\end{array}
\end{equation}
\begin{equation}
=   \tr\left({ {\mathbf{H}}_w {\mathbf{R}}_{N_t}^{\frac{1}{2}}
\mathbf{V}_b \mathbf{V}_b^H {\mathbf{R}}_{N_t }^{\frac{1}{2}} {\mathbf{H}}_w^H {\mathbf{R}}_{N_e }^{\frac{1}{2}}  \mathbf{e}_j\mathbf{e}_i^H\mathbf{R}}_{N_e }^{\frac{1}{2}} \right)
\end{equation}
\begin{equation}\label{eq:asy_cov}
\hspace{-1cm} \mathop  \to \limits^{N_t  \to \infty }  \tr({\mathbf{R}}_{N_t }^{\frac{1}{2}}
\mathbf{V}_b \mathbf{V}_b^H {\mathbf{R}}_{N_t}^{\frac{1}{2}} ) \tr\left({\mathbf{R}}_{N_e}^{\frac{1}{2}}  \mathbf{e}_j\mathbf{e}_i^H {\mathbf{R}}_{N_e}^{\frac{1}{2}}\right)
\end{equation}
\begin{equation}
\hspace{-3.2cm} =  \tr\left(\mathbf{V}_b \mathbf{V}_b^H \mathbf{R}_{N_t}\right) \mathbf{e}_i^H {\mathbf{R}}_{N_e } \mathbf{e}_j
\end{equation}
\end{subequations}
where (\ref{eq:asy_cov}) is based on \cite[Eq. (102)]{Wen2013TIT}.  As a result, we obtain
\begin{align}\label{eq:cov_jamming_2}
& \frac{{{P}_{\rm AN}}}{(N_{t} - N_r)} \mathbf{H}_{ea} \mathbf{V}_b \mathbf{V}_b^H \mathbf{H}_{ea} ^H
 +  \sigma_e^2 \mathbf{I}_{N_e}    \mathop  \to \limits^{N_t  \to \infty } \nonumber \\
&  \frac{{{P}_{\rm AN}}}{(N_{t} - N_r)} \tr\left(\mathbf{V}_b \mathbf{V}_b^H \mathbf{R}_{N_t}\right)
 \mathbf{R}_{N_e} + \sigma_e^2 \mathbf{I}_{N_e} = \mathbf{W}.
\end{align}
By whitening the noise with $\mathbf{W}^{-1/2}$ and following similar steps as in Appendix \ref{Proof_PG-GSVD_erg},
$R_{\rm{eve,\,u}}$ in (\ref{rate_erg_lower_The}) becomes
\begin{align}\label{eq:rate_jamming}
& R_{\rm{eve,\,u}}  = N_t \log M - \frac{1}{M^{N_s} }  \sum\limits_{s = 1}^{S} \sum\limits_{p_s = 1}^{M^{N_s}}  \log  \sum\limits_{q_s = 1}^{M^{N_s}}
\exp\left( \right. \nonumber \\
& \left. - {\tr({\mathbf{R}}_{N_e} \mathbf{W}^{-1})}  \mathbf{b}_{s,p_s q_s}^H \mathbf{V}_{s,\rm{erg}}^H \mathbf P_{s,\rm{erg}}^{\frac{1}{2}}
\boldsymbol{\hat{\Sigma}}_{s}^2 \mathbf P_{s,\rm{erg}}^{\frac{1}{2}} \mathbf{V}_{s,\rm{erg}} \mathbf{b}_{s,p_s q_s} \right).
\end{align}

\section{Numerical Results}
We set $\sigma_b = \sigma_e$ and define ${\rm SNR} =  P/(N_r \sigma_b^2)$. Furthermore, we use $N_t \times N_r \times N_e$
to denote the simulated wiretap channel.

\subsection{Scenarios with Instantaneous CSI of the Eavesdropper}
\begin{figure}[!t]
\centering
\includegraphics[width=0.5\textwidth]{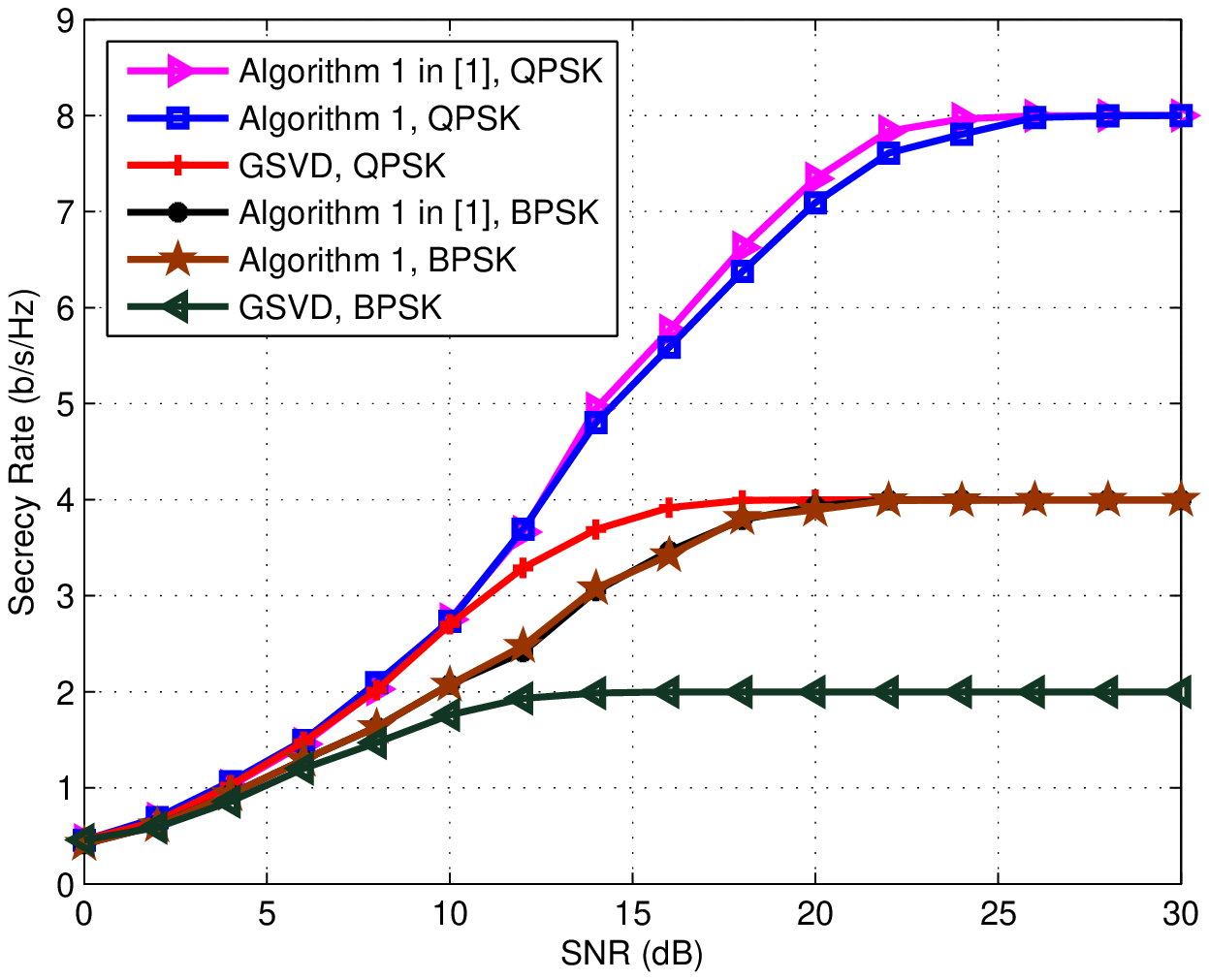}
\vspace{-10pt}
 \captionstyle{flushleft}
 \caption{Secrecy rate versus SNR for the $4\times3\times2$ wiretap channel for different precoder designs and different modulation schemes  for $N_{s} = 2$.}
\label{wiretap_421}
\end{figure}

\begin{figure}[!t]
\centering
\includegraphics[width=0.5\textwidth]{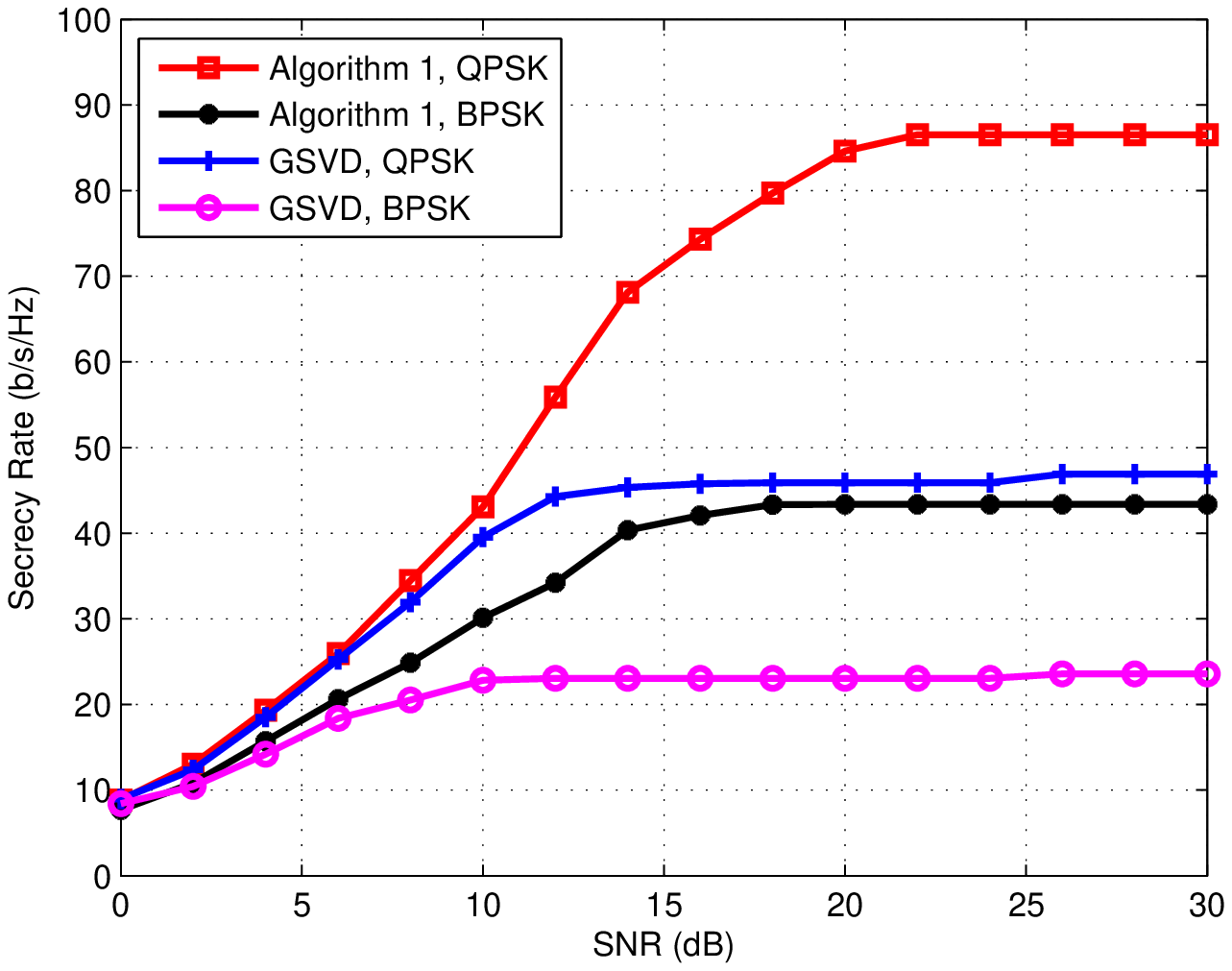}
 \vspace{-10pt}
 \captionstyle{flushleft}
\caption{Secrecy rate versus SNR for the $64\times48\times48$ wiretap channel for different precoder designs and different modulation schemes  for $N_{s} = 2$.}
\label{wiretap_64_48_48}
\end{figure}

In this subsection, the elements of $\mathbf{H}_{ba}$
and $\mathbf{H}_{ea}$ are generated independently and randomly.
Tables I and II compare the computational complexities of the different schemes
for the systems considered in Figures \ref{wiretap_421} and \ref{wiretap_64_48_48}, respectively.

\begin{table}[!t]
\centering
\caption{Number of additions required for calculating the mutual information and the MSE matrix for the system considered in Figure \ref{wiretap_421}.}
\vspace*{1.5mm}
\begin{tabular}{|c|c|c|}
\hline
$4\times3\times2$   & BPSK &  QPSK     \\ \hline
 GSVD &  8    &   16 \\ \hline
Algorithm 1 &  32   &   512 \\  \hline
 Algorithm 1 in \cite{Wu2012TVT} &   256   &   65536     \\ \hline
\end{tabular}
\end{table}

\begin{table}[!t]
\centering
\caption{Number of additions required for calculating the mutual information and the MSE matrix for the system considered in Figure \ref{wiretap_64_48_48}.}
\vspace*{1.5mm}
\begin{tabular}{|c|c|c|}
\hline
$64\times48\times48$   & BPSK &  QPSK     \\ \hline
 GSVD &  128    &   256 \\ \hline
Algorithm 1 &  512   &   8192 \\  \hline
 Algorithm 1 in \cite{Wu2012TVT} &  3.04e+038     &  1.15e+077    \\ \hline
\end{tabular}
\end{table}

Figure \ref{wiretap_421} plots the secrecy rate for the $4\times3\times2$ wiretap channel
for different precoder designs and different modulation schemes for $N_s =2$.
 We observe from Figure \ref{wiretap_421} that Algorithm 1
achieves a similar performance as the precoder design in \cite{Wu2012TVT} but with orders of magnitude lower computational complexity
as indicated in Table I.
Both designs achieve the maximal rate $ N_t \log_2 M$ b/s/Hz in the high SNR regime as indicated by Theorem \ref{PG-GSVD_high}.
In contrast, the GSVD design yields an obvious rate loss in the high SNR regime.
For the channels of Bob and Eve, we have $D_{b,1} = 0.57$ and $D_{e,1} = 0.81$, respectively.
As explained  in Example 1, the GSVD design sets $p_1 = p_2 =0$ in this case.
Therefore, the GSVD design suffers from a $2 \log_2 M$ b/s/Hz rate loss in the high SNR regime as shown in Figure  \ref{wiretap_421}.

In Figure \ref{wiretap_64_48_48}, we show the secrecy rate for the $64\times48\times48$
wiretap channel
for different precoder designs and different modulation schemes for $N_s =2$.
As indicated in Table II,
the computational complexity of the precoder design in \cite{Wu2012TVT} is prohibitive in this case and no results can
be shown. We observe that
the secrecy rate of the GSVD design is lower than the upper bound given in Theorem \ref{GSVD_loss}.
This is because for the GSVD design, as indicated in \cite[Eq. (12)]{Bashar2012TCom}, only the non-zero subchannels of Bob which are
stronger than the corresponding subchannels
of Eve can be used for transmission.  The $b_i$, $i = 1,\ldots,s$, in (\ref{eq:sigma_ba}) are in ascending order while the $e_i$, $i = 1,\ldots,s$,
in (\ref{eq:sigma_ea}) are in descending order.
Therefore, a large proportion of Bob's non-zero subchannels may be abandoned by the GSVD design for large-scale MIMO channels.
As a result, Algorithm 1 achieves significantly higher secrecy rates than the GSVD design.

\begin{figure}[!t]
\centering
\includegraphics[width=0.5\textwidth]{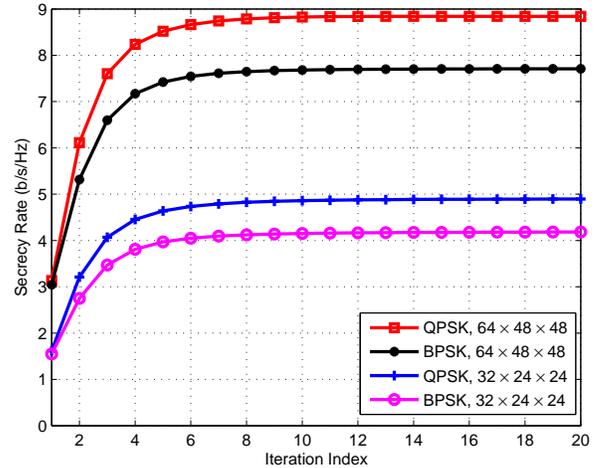}
 \vspace{-10pt}
 \captionstyle{flushleft}
\caption{Secrecy rate versus iteration index for different wiretap channels and different
modulation schemes  for Algorithm 1, $N_{s} = 2$, and ${\rm SNR} = 0$ dB.}
\label{Convergence}
\end{figure}

Figure \ref{Convergence} illustrates the convergence behavior of  Algorithm 1 for different wiretap channels and different
modulation schemes for $N_s =2$ and ${\rm SNR} = 0$ dB. Figure \ref{Convergence} shows
the secrecy rate in each iteration. We observe
that in all considered cases, Algorithm 1 converges within
a few iterations.

\begin{figure}[!t]
\centering
\includegraphics[width=0.5\textwidth]{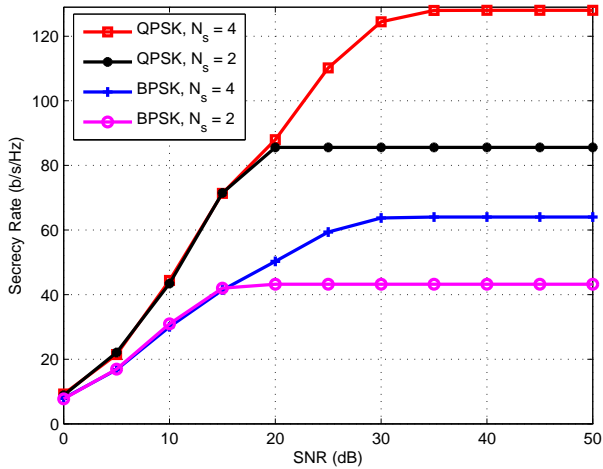}
 \vspace{-10pt}
 \captionstyle{flushleft}
\caption{Secrecy rate for Algorithm 1 versus SNR for the $64\times48\times48$ wiretap channel for different modulation schemes  and different $N_{s}$.}
\label{wiretap_64_48_48_N4}
\end{figure}

Figure \ref{wiretap_64_48_48_N4} plots the secrecy rate for Algorithm 1 for the $64\times48\times48$ wiretap channel
for different modulation schemes and different $N_s$.
For the $64\times48\times48$ wiretap channel,
$k - N_2$ is equal to $16$.
As indicated by Theorem \ref{PG-GSVD_high}, when we set $N_s = 4$,
Algorithm  1 can achieve the maximal rate $N_t \log_2 M$ b/s/Hz in the high
SNR regime. This is validated in Figure \ref{wiretap_64_48_48_N4}.

\subsection{Scenarios with Statistical CSI of the Eavesdropper}
\begin{figure}[!t]
\centering
\includegraphics[width=0.5\textwidth]{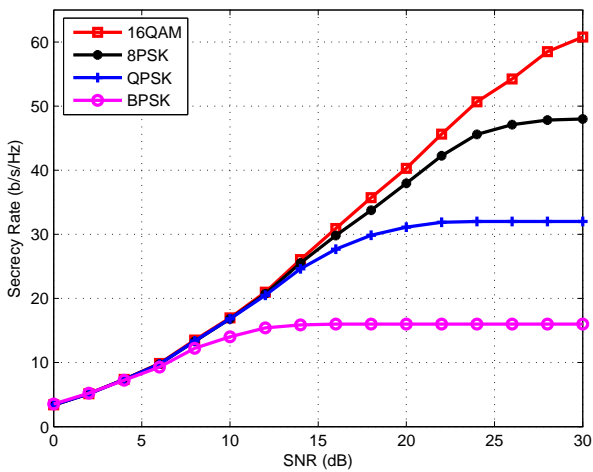}
\vspace{-10pt}
 \captionstyle{flushleft}
\caption{Secrecy rate for Algorithm 2 versus SNR for the $32\times32\times32$ wiretap channel for different modulation schemes and $N_{s} = 2$.}
\label{wiretap_32_32_32_sta_2}
\end{figure}

In Figure \ref{wiretap_32_32_32_sta_2}, we show the secrecy rate for
 Algorithm 2 for the $32\times32\times32$ wiretap channel
 for $N_s = 2$ and different modulation schemes.
{We set $\mathbf{\tilde{R}}_{N_b} = \mathbf{I}_{N_b}$ and $\mathbf{R}_{N_e} = \mathbf{I}_{N_e}$.
Also, we generate $\mathbf{\tilde{R}}_{{N}_t}$ and $\mathbf{R}_{N_t}$} based on \cite[Eq. (3.14)]{Cho2010}, where
the truncated Laplacian distribution is used to model
the channel power angle spectrum \cite{Cho2010}.
The mean  angle of arrival (AoA)  is generated randomly and the AoA
 interval is $\mathcal{A} = [-\pi/6,\pi/6]$.
The angular spread is set to be $\pi/2$.
{We generate one channel realization for the intended receiver's channel
based on (\ref{Hba_model}). Then, we evaluate the achievable ergodic
secrecy rate based on (\ref{sec_rate_erg_lower_The}).}
We observe that in the low-to-medium SNR regime, Algorithm 2
achieves nearly the same secrecy rate for all considered modulations. Furthermore, in the medium-to-high SNR regime, Algorithm 2
achieves a good secrecy rate performance for each modulation scheme.

\begin{figure}[!t]
\centering
\includegraphics[width=0.5\textwidth]{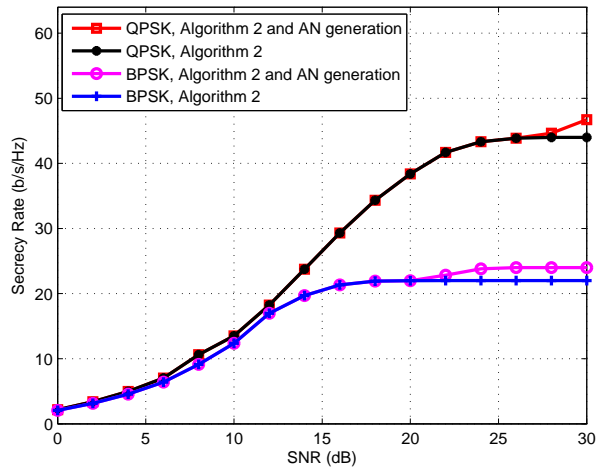}
\vspace{-10pt}
 \captionstyle{flushleft}
\caption{Secrecy rate for Algorithm 2 versus SNR for the $32\times24\times24$ wiretap channel for different modulation schemes
for $N_s =2$, and an urban scenario.}
\label{Fig_32_SCM_urban}
\end{figure}

\begin{figure}[!t]
\centering
\includegraphics[width=0.5\textwidth]{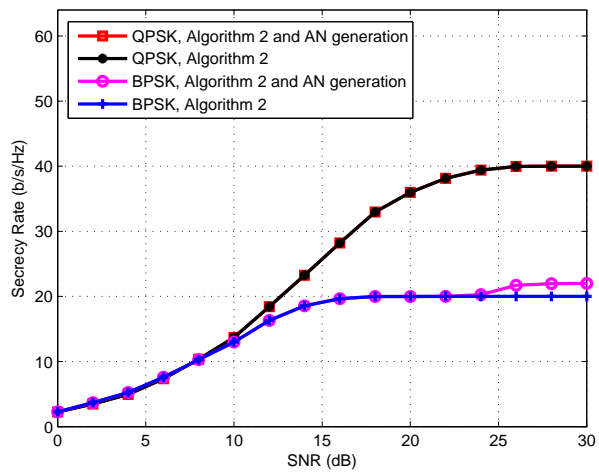}
\vspace{-10pt}
 \captionstyle{flushleft}
\caption{Secrecy rate for Algorithm 2 versus SNR for the $32\times24\times24$ wiretap channel for different modulation schemes for $N_s =2$ and a suburban scenario.}
\label{Fig_32_SCM_suburban}
\end{figure}

Figure \ref{Fig_32_SCM_urban} and Figure \ref{Fig_32_SCM_suburban} show the secrecy rate
for Algorithm 2 for the $32\times24\times24$
wiretap channel
for different modulation schemes, $N_s = 2$, and with/without artificial noise generation
for urban and suburban scenarios, respectively.
{For these figures, we adopted the 3GPP SCM \cite{Salo2005} for the urban scenario, half-wavelengh antenna spacing at transmitter and receiver, respectively, a velocity of $36$ km/h, and $6$ paths.
For the intended receiver's channel, we generate one channel realization based on the SCM model
and use it in (\ref{rate_erg}).
For the eavesdropper's channel, we generate $L = 1000$ channel realizations
$\mathbf{H}_l$, $l = 1,...,L$, based on the SCM model.
According to the Kronecker fading MIMO channel model in (\ref{Hea_model}), we can
estimate $\mathbf{R}_{N_{t}} =  \frac{1}{L}\sum\nolimits_{l=1}^{L}\mathbf{H}_l^H \mathbf{H}_l $
and $\mathbf{R}_{N_e} = \frac{1}{L} \sum\nolimits_{l=1}^{L}\mathbf{H}_l \mathbf{H}_l^H$ from these channel realizations.
For the precoder design, we substitute the obtained  $\mathbf{R}_{N_t}$ and $\mathbf{R}_{N_e}$ into the
 achievable ergodic secrecy rate expression in (\ref{sec_rate_erg_lower_The}).}
Figure \ref{Fig_32_SCM_urban} and Figure \ref{Fig_32_SCM_suburban}
indicate that Algorithm 2
achieves good secrecy rate performance for both urban
and suburban scenarios. Also, AN generation is beneficial
and achieves a secrecy rate gain in the high SNR regime.

{
We note that in the low SNR regime, Proposition 2 in [1] proves that the optimal transmission policy
for MIMO wiretap channels with finite alphabet inputs is beamforming.  Therefore, in the low SNR regime, the optimal performance of the complete search design in [1] and the proposed low complexity design  should be the same
since only one subchannel is used for transmission regardless of the value of
$N_s$. Moreover, it is also proved in [1] that for low SNR the optimal secrecy rate is independent of the constellation size.
Therefore, as indicated in Figure 5, in the low-to-moderate SNR regime, the secrecy rates of
the proposed low complexity precoder are virtually the same for QPSK, 8PSK, and 16QAM modulations.
In the high SNR regime, Theorem 3 in this paper indicates that as long as $ (k - N_2) N_s \neq N_t$
holds, we can always find a low complexity precoder design to achieve the saturation secrecy rate regardless of
the constellation size.  In the moderate-to-high SNR regime, we expect the performance gap between
the proposed low complexity design and the complete search design in [1]  to increase to some extent with the
constellation size. However, we note that even for the case of $N_t = 4$, cf.
Table I, it will be difficult to simulate the complete search design in [1] for 16QAM modulation
since the computational complexity scales linearly with $16^8$ for each iteration in Algorithm 1 in [1].
However, the proposed low complexity precoder design can be efficiently
implemented even for the case of $N_t = 32$ and 16QAM modulation,  as shown
in Figure 5.
}

\section{Conclusion}
In this paper, we have investigated the linear precoder design
for large-scale MIMOME wiretap channels with finite alphabet inputs.
For the case where the transmitter has  instantaneous CSI of the eavesdropper,
we derived an upper bound on the secrecy rate
for the GSVD design in the high SNR regime. The derived expression reveals that
the GSVD design may lead to a serious performance loss. Motivated by this,
we proposed a novel PG-GSVD design to overcome
the negative properties
of the GSVD design while retaining its low computational
complexity for large-scale MIMO systems.
We further extended
the PG-GSVD design to the case where only statistical CSI of the eavesdropper is available
at the transmitter.
For massive MIMO channels with strong transmit correlation,
we proved that the proposed PG-GSVD design
with statistical CSI of the eavesdropper can achieve
the maximal secrecy rate for finite alphabet inputs
for the MIMOME wiretap channel in the high SNR regime.
For massive MIMO channels with weak transmit correlation,
we proposed an AN generation scheme to improve the
secrecy rate  in the high SNR regime.
Simulation results indicated that the proposed designs perform well  in large-scale MIMOME
wiretap channels and achieve substantial secrecy rate gains
compared to the GSVD design for finite alphabet inputs
while requiring a substantially lower computational
complexity compared to the existing precoder design in \cite{Wu2012TVT}.
{Possible extensions of the proposed designs include the consideration of imperfect CSI of the intended user channel,
multiuser settings, and the multi-cell scenarios with pilot contamination.}

\appendices
\section{Proof of Theorem \ref{GSVD_loss}} \label{GSVD_loss_proof}
Based on (\ref{eq:sigma_ba}) and (\ref{eq:bob_precoded}), $I \left( {{\mathbf{y}}_b ;{\mathbf{x}}_a} \right)$ in (\ref{rate}) for the GSVD design becomes
\begin{equation} \label{rate_b_GSVD}
I\left( {{{\bf{y}}_b};{{\bf{x}}_a}} \right) = \sum\limits_{i = 1}^s {I\left( {b_i^2{p_{k - r - s + i}}} \right)}  + \sum\limits_{i = 1}^r {I\left( {{p_{k - s + i}}} \right)}
\end{equation}
where $I(\gamma) =  I(x; \sqrt{\gamma} x + n)$. Therefore, for $P \rightarrow \infty$, we obtain
\begin{equation} \label{rate_b_GSVD}
\mathop {\lim }\limits_{P \to \infty }  I\left( {{{\bf{y}}_b};{{\bf{x}}_a}} \right) \leq (s + r) \log_2 (M).
\end{equation}
According to  Inclusion--Exclusion Principle \cite{Andrews1971book}, we know
\begin{align} \label{Dim_set}
{\rm dim}\left(\mathcal{S}_{ba}\right) +  {\rm dim}\left(\mathcal{S}_{be}\right) = {\rm dim} \left(\mathcal{S}_{ba} \cup \mathcal{S}_{be}  \right) -
{\rm dim} \left(\mathcal{S}_{ba} \cap \mathcal{S}_{be}\right).
\end{align}
For the subspaces $\mathcal{S}_{ba}$ and $\mathcal{S}_{be}$, we have
\begin{subequations} \label{S_set_total}
\begin{equation} \label{S_set}
\begin{array}{l}
 \mathcal{S}_{ba} \cap \mathcal{S}_{be}  \\
 \hspace{-0.3cm} = \left({\rm null}\left(\mathbf{H}_{ba}\right)^{\bot}   \cap  {\rm null}\left(\mathbf{H}_{ea}\right)\right) \cap
\left( {\rm null}\left(\mathbf{H}_{ba}\right)^{\bot} \cap {\rm null}\left(\mathbf{H}_{ea}\right)^{\bot}\right)
\end{array}
\end{equation}
\begin{equation}\label{eq:Set_2}
 \hspace{-0.1cm} = \left({\rm null}\left(\mathbf{H}_{ba}\right)^{\bot}   \cap  {\rm null}\left(\mathbf{H}_{ea}\right)\right)  \cap
\left(  {\rm null}\left(\mathbf{H}_{ea}\right)^{\bot} \cap{\rm null}\left(\mathbf{H}_{ba}\right)^{\bot} \right)
\end{equation}
\begin{equation}\label{eq:Set_3}
\hspace{-0.1cm} =  {\rm null}\left(\mathbf{H}_{ba}\right)^{\bot}  \cap  \left(\left( {\rm null}\left(\mathbf{H}_{ea}\right)\right) \cap  {\rm null}\left(\mathbf{H}_{ea}\right)^{\bot}    \right)  \cap {\rm null}\left(\mathbf{H}_{ba}\right)^{\bot}
\end{equation}
\begin{equation}\label{eq:Set_4}
\hspace{-8.4cm}  = \varnothing
\end{equation}
\end{subequations}
where (\ref{eq:Set_2}) and (\ref{eq:Set_3}) are obtained based on the properties of intersections \cite{Halmos1960book}.

Also, we have
\begin{subequations} \label{S_set_cup}
\begin{equation}\label{S_set_cup_1}
\begin{array}{l}
 \mathcal{S}_{ba} \cup \mathcal{S}_{be}  \\
 \hspace{-0.3cm} =  \left({\rm null}\left(\mathbf{H}_{ba}\right)^{\bot}  \cap  {\rm null}\left(\mathbf{H}_{ea}\right)\right)  \cup
\left( {\rm null}\left(\mathbf{H}_{ba}\right)^{\bot} \!\cap\! {\rm null}\left(\mathbf{H}_{ea}\right)^{\bot}\right)  \\
\end{array}
\end{equation}
\begin{equation}\label{S_set_cup_2}
\hspace{-2.5cm} = {\rm null}\left(\mathbf{H}_{ba}\right)^{\bot} \cap \left(  {\rm null}\left(\mathbf{H}_{ea}\right) \cup {\rm null}\left(\mathbf{H}_{ea}\right)^{\bot} \right) \end{equation}
\begin{equation}\label{S_set_cup_3}
\hspace{-7cm} =  {\rm null}\left(\mathbf{H}_{ba}\right)^{\bot}
\end{equation}
\end{subequations}
where (\ref{S_set_cup_2}) and (\ref{S_set_cup_3})  are obtained based on the Distributive Law of sets \cite{Halmos1960book}
and the Rank--Nullity Theorem \cite{Banerjee2014book}, respectively.

From (\ref{Dim_set})--(\ref{S_set_cup}), we obtain
\begin{align} \label{Dim_set_sr}
s + r = {\rm dim}\left(\mathcal{S}_{ba}\right) +  {\rm dim}\left(\mathcal{S}_{be}\right) = {\rm dim}\left( {\rm null}\left(\mathbf{H}_{ba}\right)^{\bot}\right).
\end{align}

Assuming $\mathbf{v}_i \in \mathbb{C}^{N_t \times 1}$ and $\mathbf{u}_j \in \mathbb{C}^{N_r \times 1}$
are the $N_t$ left and $N_r$  right singular vectors of $\mathbf{H}_{ba}$,
respectively, $i = 1,\ldots,N_t$,  $j = 1,\ldots,N_r$, $\mathbf{H}_{ba}$ can be written as
\begin{align} \label{Dim_set_1}
\mathbf{H}_{ba} = \sum\limits_{i = 1}^{{N_1}} {{\lambda _i}{{\bf{u}}_i}{\bf{v}}_i^H}
\end{align}
where ${\lambda _i}$ is the singular value of $\mathbf{H}_{ea}$. For $N_1 < N_t$,
we have
\begin{align} \label{Dim_set_2}
{\rm null}\left(\mathbf{H}_{ba}\right) =  \sum\limits_{i = N_1 + 1}^{{N_t}} {{\omega_i}{{\bf{v}}_i}{\bf{v}}_i^H}
\end{align}
where ${\omega_i}$ denotes an arbitrary non-zero complex value, $i = 1,\ldots,N_t$.
Based on the property of the orthogonal complement of a subspace \cite{Halmos1974book}, we obtain
\begin{subequations}
\begin{equation} \label{Dim_set_3}
\hspace{-2.6cm} \left( {\rm null}\left(\mathbf{H}_{ba}\right)^{\bot}\right)  =  \left(\sum\limits_{i = N_1 + 1}^{{N_t}} {{\omega_i}{{\bf{v}}_i}{\bf{v}}_i^H}\right)^{\bot}
\end{equation}
\begin{equation}
=  {\rm null} \left(\sum\limits_{i = N_1 + 1}^{{N_t}} {{\omega_i}{{\bf{v}}_i}{\bf{v}}_i^H}\right)
\end{equation}
  \begin{equation}
\hspace{-1.6cm} =  \sum\limits_{i = 1}^{{N_1}} {{\omega_i}{{\bf{v}}_i}{\bf{v}}_i^H}.
\end{equation}
\end{subequations}

Therefore, we have
 \begin{align} \label{Dim_set_4}
 {\rm dim}\left( {\rm null}\left(\mathbf{H}_{ba}\right)^{\bot}\right) = N_1.
\end{align}
For $N_1 = N_t$,  ${\rm null}\left(\mathbf{H}_{ba}\right) = \varnothing$, and we obtain
 \begin{align} \label{Dim_set_5}
 {\rm dim}\left( {\rm null}\left(\mathbf{H}_{ba}\right)^{\bot}\right) = N_t.
\end{align}
Combining (\ref{rate}), (\ref{rate_b_GSVD}), (\ref{Dim_set_sr}), (\ref{Dim_set_4}), and (\ref{Dim_set_5}) completes the proof.

\section{Proof of Theorem \ref{PG-GSVD_high}} \label{Proof_PG-GSVD_high}
The key idea of achieving the maximal rate $N_t \log M$ b/s/Hz in the high SNR regime
is to guarantee that all $N_t$ signals can be received by Bob but not by
Eve. To achieve this, $N_s$ signals are combined into a group and transmitted along
the subchannels $\mathbf{R}_r$ in (\ref{eq:sigma_ba_hat}). As a result, we need to analyze
the dimension of $\mathcal{S}_{ba}$.

Based on the Inclusion--Exclusion Principle \cite{Andrews1971book}, we have
\begin{align} \label{Dim_set_2}
{\rm dim}\left(\mathcal{S}_{ba}\right) +  {\rm dim}\left(\mathcal{S}_{n}\right) = {\rm dim} \left(\mathcal{S}_{ba} \cup \mathcal{S}_{n}  \right) -
{\rm dim} \left(\mathcal{S}_{ba} \cap \mathcal{S}_{n}\right).
\end{align}

Following similar steps as in (\ref{S_set_total}) and (\ref{S_set_cup}), we obtain
\begin{subequations}
\begin{equation} \label{S_set_high}
\begin{array}{l}
 \mathcal{S}_{ba} \cap \mathcal{S}_{n} \\
  = \left({\rm null}\left(\mathbf{H}_{ba}\right)^{\bot}  \cap {\rm null}\left(\mathbf{H}_{ea}\right)\right)  \cap
\left( {\rm null}\left(\mathbf{H}_{ba}\right) \cap {\rm null}\left(\mathbf{H}_{ea}\right)\right)
\end{array}
\end{equation}
\begin{equation} \label{eq:Set_2_high}
 = \left(   {\rm null}\left(\mathbf{H}_{ea}\right)  \cap   {\rm null}\left(\mathbf{H}_{ba}\right)^{\bot} \right)  \cap
\left( {\rm null}\left(\mathbf{H}_{ba}\right) \cap {\rm null}\left(\mathbf{H}_{ea}\right)\right)
\end{equation}
\begin{equation} \label{eq:Set_3_high}
\hspace{-0.2cm} =  {\rm null}\left(\mathbf{H}_{ea}\right) \! \cap \! \left(\left( {\rm null}\left(\mathbf{H}_{ba}\right)\right)^{\bot}  \cap {\rm null}\left(\mathbf{H}_{ba}\right)    \right) \! \cap \! {\rm null}\left(\mathbf{H}_{ea}\right)
\end{equation}
\begin{equation} \label{eq:Set_4_high}
\hspace{-7.9cm} = \varnothing
\end{equation}
\end{subequations}
and
\begin{subequations}
\begin{equation} \label{S_set_cup_1_high}
\begin{array}{l}
 \mathcal{S}_{ba} \cup \mathcal{S}_{n} \\
  = \left({\rm null}\left(\mathbf{H}_{ba}\right)^{\bot}   \cap  {\rm null}\left(\mathbf{H}_{ea}\right)\right)  \cup
\left( {\rm null}\left(\mathbf{H}_{ba}\right) \cap {\rm null}\left(\mathbf{H}_{ea}\right) \right)
\end{array}
\end{equation}
\begin{equation} \label{S_set_cup_2_high}
\hspace{-2.2cm}   = {\rm null}\left(\mathbf{H}_{ea}\right) \cap \left(  {\rm null}\left(\mathbf{H}_{ba}\right)^{\bot} \cup {\rm null}\left(\mathbf{H}_{ea}\right) \right)
\end{equation}
\begin{equation}\label{S_set_cup_3_high}
\hspace{-6.5cm} =  {\rm null}\left(\mathbf{H}_{ea}\right).
\end{equation}
\end{subequations}
Since ${\rm rank} \left(\mathbf{H}_{ea}\right) = N_2$,  we have ${\rm dim} \left({\rm null}\left(\mathbf{H}_{ea}\right)\right) = N_t - N_2$.
Then, based on (\ref{Dim_set_2}), (\ref{eq:Set_4_high}), (\ref{S_set_cup_3_high}),  we obtain
\begin{align} \label{Dim_set_3}
r + N_t - k = N_t - N_2.
\end{align}
From (\ref{Dim_set_3}), we know $r = k - N_2$.

When $(k - N_2) N_s \geq N_t$, we design the PG-GSVD precoder in (\ref{eq:precoding_matrix_gsvd}) as follows. We set
\begin{equation}\label{eq:P_high}
\mathbf{P} = \kbordermatrix {~		   &  k -r-s	 & s 		& r 	& N_t - k	\cr
										k - r - s & \mathbf{0} 	& \mathbf{0}  		& \mathbf{0} & \mathbf{0}	\cr
										s                 & \mathbf{0}  	&  \mathbf{0} 	& \mathbf{0} & \mathbf{0}	\cr
										r                 & \mathbf{0}  	& \mathbf{0}      	& {\rm diag}\left(p_1,\ldots p_r\right) 	& \mathbf{0}  \cr
                                       N_t -k                 & \mathbf{0}  	& \mathbf{0}      	& \mathbf{0} 	& \mathbf{0}  \cr  }.
\end{equation}
Also, we select a pairing scheme $\left\{\ell_{1},\ldots,\ell_{N_{t}}\right\}$ in (\ref{eq:P_pair}) satisfying
\begin{equation}\label{eq:pair_high}
\left[\mathbf{P}_s\right]_{ii} = \\
 \left\{ \begin{array}{l}
   0  \quad \quad \quad {\rm if} \  1 \leq i  \leq N_s -1 \\
\frac{p_j}{\omega_{k - r + j}} \ \, {\rm if}  \  i = N_s
 \end{array} \right.
\end{equation}
for $s = 1,\ldots, S$, $i = 1,\ldots, N_s$, and $j = 1,\ldots,r$.

Based on the design in (\ref{eq:P_high}) and (\ref{eq:pair_high}), in the high SNR regime, we have
\begin{align}
 I\left( {{\bf{y}}_{b,s}};\mathbf{x}_{s} \right) & \mathop  \to \limits^{P \to \infty } N_s \log M \label{eq:bob_high}\\
  I\left( {{\bf{y}}_{e,s}};\mathbf{x}_{s} \right)& = 0.  \label{eq:eve_high}
\end{align}

Substituting (\ref{eq:bob_high}) and (\ref{eq:eve_high}) into (\ref{eq:I_pair}) completes the proof.

\section{Proof of Theorem \ref{PG-GSVD_erg}} \label{Proof_PG-GSVD_erg}
By setting $\mathbf{G}_{\rm{erg}} = \mathbf{U}_{a,\rm{erg}} \mathbf{A}_{\rm{erg}} \mathbf P_{\rm{erg}}^{\frac{1}{2}} \mathbf{V}_{\rm{erg}}$,
we know  $I \left( {{\mathbf{y}}_b ;{\mathbf{x}}_a} \right)$
in (\ref{rate_erg}) can be written as
\begin{equation}\label{eq:I_pair_erg}
I \left( {{\mathbf{y}}_b ;{\mathbf{x}}_a} \right) = \sum\limits_{s = 1}^S I\left( {{\bf{y}}_{b,s,\rm{erg}}};\mathbf{x}_{s,\rm{erg}} \right).
\end{equation}

Next, we consider $E\left[I \left( {{\mathbf{y}}_{e} ;{\mathbf{x}}_a} \right)\right]$.
 In the expression of $I \left( {{\mathbf{y}}_{e} ;{\mathbf{x}}_a} \right)$  in
\cite[Eq. (12)]{Wu2012TVT}, $\log\exp(\sum\limits x_k)$ is a convex function.
Therefore, by applying Jensen's inequality to $I \left( {{\mathbf{y}}_{e} ;{\mathbf{x}}_a} \right)$,
we obtain an upper bound on  $E\left[I \left( {{\mathbf{y}}_{e} ;{\mathbf{x}}_a} \right)\right]$
\begin{align}\label{rate_erg_lower_1}
& E\left[I \left( {{\mathbf{y}}_{e} ;{\mathbf{x}}_a} \right)\right] \leq R_{\rm{eve,\,u}} = N_t \log M  \nonumber \\
 & -   \frac{1}{M^{N_t}}  \sum\limits_{p = 1}^{M^{N_t}} \log \sum\limits_{q = 1}^{M^{N_t}} \exp\left(- \frac{\tr({\mathbf{R}}_{N_e})}{\sigma_e^2}
\mathbf{b}_{pq}^H \mathbf{G}^H {\mathbf{R}}_{N_t } \mathbf{G} \mathbf{b}_{pq}  \right)
\end{align}

Substituting $\mathbf{G} =\mathbf{G}_{\rm{erg}}$, $\mathbf{R}_{N_t} = \mathbf{T}^H \mathbf{T}$,
and (\ref{eq:sigma_ea_erg}) in (\ref{rate_erg_lower_1}), we have
\begin{align}\label{rate_erg_lower_gsvd}
& R_{\rm{eve,\,u}}  = N_t \log M - \frac{1}{M^{N_t} }\sum\limits_{p = 1}^{M^{N_t}} \log \nonumber \\
&    \sum\limits_{q = 1}^{M^{N_t}}
\exp\left(- \frac{\tr({\mathbf{R}}_{N_e})}{\sigma_e^2} \mathbf{b}_{pq}^H \mathbf{V}_{\rm{erg}}^H \mathbf P_{\rm{erg}}^{\frac{1}{2}}
\boldsymbol{\tilde{\Sigma}}_{ea,\rm{erg}}^{2} \mathbf P_{\rm{erg}}^{\frac{1}{2}} \mathbf{V}_{\rm{erg}} \mathbf{b}_{pq} \right).
\end{align}

Considering the structure of $\mathbf P_{\rm{erg}}$ and $\mathbf{V}_{\rm{erg}}$ in Section \ref{sec:PG-GSVD_sta},
$R_{\rm{eve,\,u}}$ can be further simplified as in (\ref{rate_erg_lower_2})--(\ref{rate_erg_lower_3})
at the top of the next page.

 \begin{figure*}[!ht]
\begin{subequations}
\begin{equation}\label{rate_erg_lower_2}
\begin{array}{l}
\hspace{-2.7cm} R_{\rm{eve,\,u}}  \\
\hspace{-2.7cm}  =  N_t \log M -  \frac{1}{M^{N_t} } \sum\limits_{p = 1}^{M^{N_t}} \log \sum\limits_{q = 1}^{M^{N_t}}
\exp\left( - \frac{\tr({\mathbf{R}}_{N_e})}{\sigma_e^2} \sum\limits_{s = 1}^{S} \mathbf{b}_{s,pq}^H \mathbf{V}_{s,\rm{erg}}^H \mathbf P_{s,\rm{erg}}^{\frac{1}{2}}
\boldsymbol{\hat{\Sigma}}_{s}^2 \mathbf P_{s,\rm{erg}}^{\frac{1}{2}} \mathbf{V}_{s,\rm{erg}} \mathbf{b}_{s,pq} \right)
\end{array}
\end{equation}
\begin{equation}
\hspace{-2.4cm}  =  N_t \log M -  \frac{1}{M^{N_t} } \sum\limits_{p = 1}^{M^{N_t}} \log \sum\limits_{q = 1}^{M^{N_t}}
\prod\limits_{s = 1}^{S} \exp\left( - \frac{\tr({\mathbf{R}}_{N_e})}{\sigma_e^2}  \mathbf{b}_{s,pq}^H \mathbf{V}_{s,\rm{erg}}^H \mathbf P_{s,\rm{erg}}^{\frac{1}{2}}
\boldsymbol{\hat{\Sigma}}_{s}^2 \mathbf P_{s,\rm{erg}}^{\frac{1}{2}} \mathbf{V}_{s,\rm{erg}} \mathbf{b}_{s,pq} \right)
\end{equation}
\begin{equation}
\hspace{-0.9cm}  =  N_t \log M - \frac{1} {M^{N_t} } \sum\limits_{p = 1}^{M^{N_t}} \log  \sum\limits_{q_1 = 1}^{M^{N_s}} \cdots \sum\limits_{q_{S} = 1}^{M^{N_s}}
 \prod\limits_{s = 1}^{S} \exp\left( - \frac{\tr({\mathbf{R}}_{N_e})}{\sigma_e^2}  \mathbf{b}_{s,pq_s}^H \mathbf{V}_{s,\rm{erg}}^H \mathbf P_{s,\rm{erg}}^{\frac{1}{2}}
\boldsymbol{\hat{\Sigma}}_{s}^2 \mathbf P_{s,\rm{erg}}^{\frac{1}{2}} \mathbf{V}_{s,\rm{erg}} \mathbf{b}_{s,pq_s} \right)
\end{equation}
\begin{equation}
\begin{array}{l}
\hspace{-2.8cm} =  N_t \log M - \frac{1 }{M^{N_t} } \sum\limits_{p = 1}^{M^{N_t}} \log  \sum\limits_{q_1 = 1}^{M^{N_s}} \exp\left( -
 \frac{\tr({\mathbf{R}}_{N_e})}{\sigma_e^2}  \mathbf{b}_{1,pq_1}^H
 \mathbf{V}_{1,\rm{erg}}^H \mathbf P_{1,\rm{erg}}^{\frac{1}{2}}
\boldsymbol{\hat{\Sigma}}_{1}^2 \mathbf P_{1,\rm{erg}}^{\frac{1}{2}} \mathbf{V}_{1,\rm{erg}} \mathbf{b}_{1,pq_1} \right)   \\
 \times  \cdots \times \sum\limits_{q_{S} = 1}^{M^{N_s}} \exp\left( - \frac{\tr({\mathbf{R}}_{N_e})}{\sigma_e^2}  \mathbf{b}_{S,pq_S}^H
 \mathbf{V}_{S,\rm{erg}}^H \mathbf P_{S,\rm{erg}}^{\frac{1}{2}}
\boldsymbol{\hat{\Sigma}}_{S}^2 \mathbf P_{S,\rm{erg}}^{\frac{1}{2}} \mathbf{V}_{S,\rm{erg}} \mathbf{b}_{S,pq_S} \right)
\end{array}
\end{equation}
\begin{equation}
\hspace{-2cm}  = N_t \log M - \frac{1 }{M^{N_t} }\sum\limits_{p = 1}^{M^{N_t}} \sum\limits_{s = 1}^{S} \log  \sum\limits_{q_s = 1}^{M^{N_s}} \exp\left(
  - \frac{\tr({\mathbf{R}}_{N_e})}{\sigma_e^2}  \mathbf{b}_{s,pq_s}^H
 \mathbf{V}_{s,\rm{erg}}^H \mathbf P_{s,\rm{erg}}^{\frac{1}{2}}
\boldsymbol{\hat{\Sigma}}_{s}^2 \mathbf P_{s,\rm{erg}}^{\frac{1}{2}} \mathbf{V}_{s,\rm{erg}} \mathbf{b}_{s,pq_s} \right)
\end{equation}
\begin{equation}\label{rate_erg_lower_3}
\hspace{-1.5cm} = N_t \log M - \frac{1}{M^{N_t} } \sum\limits_{s = 1}^{S} \sum\limits_{p_s = 1}^{M^{N_s}}  \log  \sum\limits_{q_s = 1}^{M^{N_s}} \exp\left(
  - \frac{\tr({\mathbf{R}}_{N_e}) }{\sigma_e^2}  \mathbf{b}_{s,p_s q_s}^H
 \mathbf{V}_{s,\rm{erg}}^H \mathbf P_{s,\rm{erg}}^{\frac{1}{2}}
\boldsymbol{\hat{\Sigma}}_{s}^2 \mathbf P_{s,\rm{erg}}^{\frac{1}{2}} \mathbf{V}_{s,\rm{erg}} \mathbf{b}_{s,p_s q_s} \right).
\end{equation}
\end{subequations}
 \hrulefill
\vspace*{4pt}
\end{figure*}

Combining (\ref{rate_erg}), (\ref{eq:I_pair_erg}), (\ref{rate_erg_lower_1}), and (\ref{rate_erg_lower_3})
completes the proof.


\end{document}